\documentclass[a4paper,twoside]{article}

\newcommand{\slpart}{\mbox{$\partial \hspace{-0.50em}/$}}
\newcommand{\slp}{\mbox{$p \hspace{-0.45em}/$}}

\usepackage{amssymb}
\usepackage{amsmath}
\usepackage{amsfonts}
\usepackage{ulem}
\usepackage[toc,page]{appendix}
\usepackage{hyperref}

\title{On the Quantization of the Higher Spin Fields}
\author{J.W.Wagenaar and T.A.Rijken}

\begin{document}
\allowdisplaybreaks \maketitle

\abstract{In this article we quantize (massive) higher spin
($1\leq j\leq2$) fields by means of Dirac's Constrained Hamilton
procedure both in the situation were they are totally free and
were they are coupled to (an) auxiliary field(s). A full
constraint analysis and quantization is presented by determining
and discussing all constraints and Lagrange multipliers and by
giving all equal times (anti) commutation relations. Also we
construct the relevant propagators. In the free case we obtain the
well-known propagators and show that they are not covariant, which
is also well known. In the coupled case we do obtain covariant
propagators (in the spin-3/2 case this requires $b=0$) and show
that they have a smooth massless limit connecting perfectly to the
massless case (with auxiliary fields). We notice that in our
system of the spin-3/2 and spin-2 case the massive propagators
coupled to conserved currents only have a smooth limit to the pure
massless spin-propagator, when there are ghosts in the massive
case.}

\section{Introduction}\label{intro1}

This article is about the quantization of higher spin ($1\leq
j\leq2$) fields and their propagators. Besides the interest in
their own, the physical interest in these various fields comes
from very different areas in (high energy) physics. The massive
spin-1 field is extremely important in the electro-weak part of
the Standard Model and in phenomenological One-Boson-Exchange
(OBE) models. Needless to mention the physical interest in the
photon.

As far as the spin-3/2 field is concerned, ever since the
pioneering work of \cite{fierz} and \cite{rarita} it has been
considered by many authors for several reasons. The spin-3/2 field
plays a significant role in low energy hadron scattering, where it
appears as a resonance. Also in supergravity (for a review see
\cite{pvn}) and superstring theory the spin-3/2 field plays an
important role, since it appears in these theories as a massless
gravitino. Besides the role it plays in the tensor-force in
OBE-models the spin-2 field mainly appears in (super-) gravity and
string theories as the massless graviton.

The quantization of such fields can roughly be divided in three
areas: free field quantization, the quantization of the system
where it is coupled to (an) auxiliary field(s) and the
quantization of an interacting field. The latter area in the
spin-3/2 case is known to have problems and inconsistencies (see
for instance \cite{sudar}, \cite{velo} and \cite{pasctim}).
Although very interesting, in this article we will focus our
attention on the first two areas.

In section \ref{ffields} we start with the quantization of the
massive, free fields. We do this for all spin cases
($j=1,\,3/2,\,2$) at the same time using Dirac's prescription
\cite{Dirac}. The inclusion of the spin-1 field case is merely
meant to demonstrate Dirac's procedure in a simple case and to
have a complete description of higher spin field quantization.

The free spin-3/2 field quantization is in the same line as in
references \cite{senjan,baaklini,pasc,pascthes}. In \cite{senjan}
the massless free spin-3/2 field was quantized in the transverse
gauge. The authors of \cite{baaklini,pasc} quantize the massive
free theory, which is also what we do. We will follow Dirac's
prescription straightforwardly by first determining all Lagrange
multipliers and constraints. Afterwards the Dirac bracket (Db) is
introduced and we calculate the equal time anti commutation (ETAC)
relations among all components of the field. In both
\cite{baaklini} and \cite{pasc,pascthes} the step to the Dirac
bracket is made earlier, without determining all Lagrange
multipliers and constraints. In \cite{baaklini} it is mentioned
that this involves "technical difficulties and much labor" and in
\cite{pasc,pascthes} the focus is on the number of constraints and
therefore not so much on their specific forms. As a result
\cite{baaklini} and \cite{pasc,pascthes} both calculate only the
ETAC relations between the spatial components of the spin-3/2
field, whereas we obtain them all.

A Dirac constraint analysis of the free spin-2 field can be found
for instance in \cite{ghal,green,baaklini1}. In these references
the massless (\cite{ghal,green}) case and massive
(\cite{baaklini1}) case is considered. We stress, however, that
our description of the quantization not only differs from
\cite{baaklini1} in the sense that the nature of one of the
obtained constraints is different, which we will discuss below,
also we obtain all constraints and Lagrange multipliers by
applying Dirac's procedure straightforward. We present a full
analysis of the constrained system. After introducing the Dirac
bracket (Db) we give all equal time commutation (ETC) relations
between the various components of the spin-2 field.

Having quantized the free theories properly we make use of a free
field expansion identity and with these ingredients we obtain the
propagators. We notice that they are not explicitly covariant, as
is mentioned for instance in \cite{weinberg} for general cases
$j\geq1$.\\

\noindent To cure this problem we are inspired by \cite{Nakan2}
and allow for auxiliary fields in the free Lagrangian in section
\ref{aux}. To be more specific we couple gauge conditions of the
massless cases to auxiliary fields and also allow for mass terms
of these auxiliary fields, with which free (gauge) parameters are
introduced. As in for instance \cite{Nakan2}, we obtain a
covariant vector field propagator, independently of the choice of
the parameter.

In the spin-3/2 case several systems of a spin-3/2 field coupled
to auxiliary fields are considered in \cite{babu,kimura,endo}. In
\cite{kimura,endo} are for several of such systems four
dimensional commutation relations obtained. In the only massive
case which the authors of \cite{endo} consider, two auxiliary
fields are introduced to couple (indirectly) to the constraint
equations \footnote{$i\partial\psi=0$ is a constraint in the sense
that it reduces the number of degrees of freedom of a general
$\psi_{\mu}$ field. It is not a constraint in the sense of Dirac,
since it is a dynamical equation.} of a spin-3/2 field. The
authors of \cite{babu} use the Lagrange multiplier \footnote{These
Lagrange multipliers are the ones used in the original sense and
are therefore different then the ones used in Dirac's formalism.}
method, where this multiplier is coupled to the covariant gauge
condition of the massless spin-3/2 field in the Rarita-Schwinger
(RS) framework (to be defined below). They notice that the
Lagrange multiplier has to be a spinor and in this sense it can
also be viewed as an auxiliary field. We follow the same line by
coupling our auxiliary field to the above mentioned gauge
condition. In \cite{babu} the quantization is performed outside
the RS framework in order to circumvent the appearance of
singularities. We remain within the RS framework and deal with
these singularities relying on Dirac's method. Therefore we stay
in line with the considerations of section \ref{ffields}. A
covariant propagator is obtained for one specific choice of the
parameter ($b=0$). This propagator is the same as the one obtained
in \cite{babu}. We notice that also in \cite{munc} a covariant
propagator is obtained, but these authors make use of two spin-1/2
fields.

Coupled systems of spin-2 and auxiliary fields were for various
reasons considered in for instance
\cite{kimura2,nakan2,babu2,gomis,klish}. In \cite{nakan2} an
auxiliary boson field is coupled to the "De Donder" gauge
condition in the Lagrangian which also contains Faddeev-Popov
ghosts. In \cite{babu2} an auxiliary field is coupled to the
divergence of the tensor field in such a way that the auxiliary
field can be viewed as a Lagrange multiplier. These authors
mention that if an other auxiliary field  is introduced, coupled
to the trace of the tensor field in order to get the other spin-2
condition, four dimensional commutation relations for the tensor
field can not be written down. We present a description in which
this is possible relying on Dirac's procedure. Also in the tensor
field case we obtain a covariant propagator, independently of the
choice of the parameter.

Having obtained all the various covariant propagators we discuss
several choices of the parameters (if possible) and the massless
limits of these propagators. We show that the propagators do not
only have a smooth massless limit but that they also connect to
the ones obtained in the massless case (including (an) auxiliary
field(s)).

When coupled to conserved currents we see that it is possible to
obtain the correct massless spin-$j$ propagators carrying only the
helicities $\lambda=\pm j_z$. This does not require a choice of
the parameter in the spin-1 case, but in the spin-3/2 and in the
spin-2 case we have to make the choices $b=0$ \footnote{This
choice we already made in order to obtain a covariant propagator.}
and $c=\pm\infty$. As far as these last two cases is concerned, it
is a different situation then taking the massive propagator,
couple it to conserved currents and putting the mass to zero as
noticed in \cite{deser} and \cite{veltman}, respectively. A
discussion on the latter matter in (anti)-de Sitter spaces can be
found in \cite{porrati,kogan,duff}. We stress however, that in the
spin-3/2 and the spin-2 case this limit is only smooth if the
massive propagator contains ghosts.

\section{Free Fields}\label{ffields}

As mentioned in the introduction we deal with the free theories in
this section. We start in section \ref{eom1} with the Lagrangians
and the equations of motion that can be deduced from them. We
explicitly quantize the theories in section \ref{quant1} and
calculate the propagators in section \ref{prop1}.

\subsection{Equations of Motion}\label{eom1}

As a starting point we take the Lagrangian for free, massive
fields ($j=1,\,3/2,\,2$). In case of the spin-3/2 there is,
according to \cite{pascthes,moldau,aurillia,nath,benmerrouche}, a
class of Lagrangians describing the particularities of a spin-3/2
field. Also in the spin-2 case several authors
(\cite{babu2,nath2,bhar,MacF}) describe a class of Lagrangians
(with one or more free parameters) which give the correct
Euler-Lagrange equations for a spin-2 field. By taking this spin-2
field to be real and symmetric from the outset only one parameter
remains
\begin{subequations}
\begin{eqnarray}
 \mathcal{L}_{1}
&=&
 -\frac{1}{2}\left(\partial_\mu A_\nu \partial^\mu A^\nu
 -\partial_\mu A_\nu\partial^\nu A^\mu\right)+\frac{1}{2}\,M_1^2A^\mu
 A_\mu\ ,\label{eqn1.1a}\\
 \mathcal{L}_{3/2,A}
&=&
 \bar{\psi}^{\mu}\left[(i\slpart-M_{3/2})g_{\mu\nu}
 +A(\gamma_{\mu}i\partial_{\nu}+\gamma_{\nu}i\partial_{\mu})
 +B\gamma_{\mu}i\slpart\gamma_{\nu}\vphantom{\frac{A}{A}}
 +CM_{3/2}\gamma_{\mu}\gamma_{\nu}\right]\psi^{\nu}\ ,\qquad\quad\label{eqn1.1b}\\
 \mathcal{L}_{2,A}
&=&
 \frac{1}{4}\,\partial^{\alpha}h^{\mu\nu}\partial_{\alpha}h_{\mu\nu}
 -\frac{1}{2}\,\partial_{\mu}h^{\mu\nu}\partial^{\alpha}h_{\alpha\nu}
 -\frac{1}{4}\,B\,\partial_{\nu}h^{\beta}_{\beta}\partial^{\nu}h^{\alpha}_{\alpha}
 -\frac{1}{2}\,A\,\partial_{\alpha}h^{\alpha\beta}\partial_{\beta}h^{\nu}_{\nu}
 \nonumber\\
&&
 -\frac{1}{4}\,M_2^2h^{\mu\nu}h_{\mu\nu}+\frac{1}{4}\,CM_2^2h^{\mu}_{\mu}h^{\nu}_{\nu}\
 ,\label{eqn1.1c}
\end{eqnarray}\label{eqn1.1}
\end{subequations}
where $B=\frac{1}{2}(3A^2+2A+1)$, $C=3A^2+3A+1$ and
$A\neq-\frac{1}{2}$, but arbitrary otherwise. We improperly
\footnote{Although the authors of \cite{rarita} mention a general
class, they expose one specific Lagrangian which would correspond
to the choice $A=-\frac{1}{3}$} refer to (\ref{eqn1.1b}) as the RS
case.

Since we do not need to be so general we choose $A=-1$ and end-up
with a particular spin-3/2 Lagrangian also used in
\cite{pvn,senjan,baaklini,pasc,pascthes,endo} and in case of the
spin-2 field we get the well-know Fierz-Pauli Lagrangian
\cite{fierz} also used in for instance
\cite{schwinger,salam,leclerc}
\begin{subequations}
\begin{eqnarray}
 \mathcal{L}_{3/2}
&=&
 -\frac{1}{2}\,\epsilon^{\mu\nu\rho\sigma}\bar{\psi}_{\mu}
 \gamma_5\gamma_{\rho}\left(\partial_{\sigma}\psi_{\nu}\right)
 +\frac{1}{2}\,\epsilon^{\mu\nu\rho\sigma}
 \left(\partial_{\sigma}\bar{\psi}_{\mu}\right)\gamma_5\gamma_{\rho}\psi_{\nu}
 -M_{3/2}\bar{\psi}_{\mu}\sigma^{\mu\nu}\psi_{\nu}
 \ ,\label{eqn1.2a}\\
 \mathcal{L}_2
&=&
 \frac{1}{4}\,\partial^{\alpha}h^{\mu\nu}\partial_{\alpha}h_{\mu\nu}
 -\frac{1}{2}\,\partial_{\mu}h^{\mu\nu}\partial^{\alpha}h_{\alpha\nu}
 -\frac{1}{4}\,\partial_{\nu}h^{\beta}_{\beta}\partial^{\nu}h^{\alpha}_{\alpha}
 +\frac{1}{2}\,\partial_{\alpha}h^{\alpha\beta}\partial_{\beta}h^{\nu}_{\nu}
 \nonumber\\
&&
 -\frac{1}{4}\,M_2^2h^{\mu\nu}h_{\mu\nu}+\frac{1}{4}\,M_2^2h^{\mu}_{\mu}h^{\nu}_{\nu}
 \ .\label{eqn1.2b}
\end{eqnarray}
\end{subequations}
Although we have picked particular Lagrangians we can always go
back to the general case by redefining the fields in the following
sense
\begin{eqnarray}
\begin{array}{ll}
 \psi'_{\mu}=O_{\mu}^{\alpha}(A)\psi_{\alpha}\ , &
 O_{\mu}^{\alpha}(A)=g_\mu^\alpha-\frac{A+1}{2}\,\gamma_\mu\gamma^\alpha\ ,\\
 h'_{\mu\nu}=O_{\mu\nu}^{\alpha\beta}(A)h_{\alpha\beta}\ , &
 O_{\mu\nu}^{\alpha\beta}(A)=\frac{1}{2}\left(g_\mu^\alpha g_\nu^\beta+g_\mu^\beta
 g_\nu^\alpha-(A+1)g_{\mu\nu}g^{\alpha\beta}\right)\ .\label{eqn1.3a}
\end{array}\label{eqn1.3b}
\end{eqnarray}
The transformation in the first line of \eqref{eqn1.3a} was also
mentioned in \cite{pascthes}. Requiring that the transformation
matrices in \eqref{eqn1.3b} are non-singular ($detO\neq0$) gives
again the constraint $A\neq-\frac{1}{2}$.

The Euler-Lagrange equations following from the free field
Lagrangians lead to the correct equations of motion (EoM)
\begin{eqnarray}
 (\Box+M_1^2)A^\mu=0\quad , & \quad\partial\cdot A=0\quad , &
 \nonumber\\*
 (i\slpart-M_{3/2})\psi_\mu=0\quad , & \quad\gamma\cdot\psi=0\quad , & \quad i\partial\cdot\psi=0\
 ,\nonumber\\*
 (\Box+M_2^2)h^{\mu\nu}=0\quad , & \quad\partial_{\mu}h^{\mu\nu}=0\quad , & \quad h^{\mu}_{\mu}=0\ .\label{eqn1.3}
\end{eqnarray}
The massless versions of the Lagrangians $\mathcal{L}_{1}$,
$\mathcal{L}_{3/2}$ and $\mathcal{L}_{2}$ \footnote{The massless
version of (\ref{eqn1.2b}) is the linearized Einstein-Hilbert
Lagrangian discussed in many textbooks as for instance
\cite{dinverno}} exhibit a gauge freedom: they are invariant under
the transformations $A^{\mu}\rightarrow{A^{\mu}}'=
A^{\mu}+\partial^{\mu}\Lambda$,
$\psi_\mu\rightarrow\psi'_\mu=\psi_\mu+\partial_\mu\epsilon$ and
${h^{\mu\nu}}\rightarrow{h^{\mu\nu}}'=
h^{\mu\nu}+\partial^{\mu}\eta^{\nu}+\partial^{\nu}\eta^{\mu}$ as
well as ${h^{\mu\nu}}\rightarrow{h^{\mu\nu}}'=
h^{\mu\nu}+\partial^{\mu}\partial^{\nu}\Lambda$, respectively.
Here, $\Lambda$, $\epsilon$ and $\eta^\mu$ are scalar, spinor and
vector fields, respectively.

In the spin-1 case a popular gauge is the Lorentz gauge
$\partial\cdot A=0$. Imposing this gauge conditions automatically
ensures the EoM $\Box A^\mu=0$ and puts the constraint
$\Box\Lambda=0$. This last constraint is used to eliminate the
residual helicity state $\lambda=0$.

A popular gauge in the spin-3/2 case  is the covariant gauge
$\gamma\cdot\psi=0$, which causes similar effects, namely the
correct EoM $i\slpart\psi=0$ and $i\partial\cdot\psi=0$ and the
constraint $i\slpart\epsilon$. Since the $\epsilon$-field is a
free spinor, it is used to transform away the helicity states
$\lambda=\pm 1/2$ of the free $\psi_\mu$ field.

Since the spin-2 Lagrangian has two symmetries, two gauge
conditions need to be imposed. The gauge conditions
$h^{\alpha}_{\alpha}=0$ and $\partial_{\alpha}h^{\alpha\beta}=0$
give the correct EoM. From the effects these gauge conditions have
on the auxiliary fields ($\Box\eta^\mu=0$, $\partial\cdot\eta=0$
and $\Box\Lambda=0$) we see that these equations describe a
massless spin-1 field and a massless spin-0 field. Therefore these
fields can be used to ensure that the tensor field $h^{\mu\nu}$
only has $\lambda=\pm 2$ helicity states.

In our case the mass terms in the Lagrangian break the gauge
symmetry. Although, the correct EoM \eqref{eqn1.3} are obtained
the freedom in the choice of the field can not be exploited to
transform away helicity states. Therefore, the massive fields
contain all helicity states, as is of course well known.

\subsection{Quantization}\label{quant1}

For the quantization of our systems we use Dirac's Hamilton method
for constrained systems \cite{Dirac}. In case of the (real) vector
and tensor fields the accompanying canonical momenta are defined
in the usual way. Since we use complex fields in case of the
spin-3/2 field we consider $\psi_\mu$ and $\psi^\dagger_\mu$ as
independent fields being elements of a Grassmann algebra. For the
definition of the accompanying canonical momenta we rely on
\cite{Hiida}. Although, the authors of \cite{Hiida} use spin-1/2
fields, the prescription for the canonical momenta does not
change. The canonical momenta are defined as
\begin{eqnarray}
 \pi^\nu_a=\frac{\partial^r\mathcal{L}}{\partial\dot{\psi}_{a,\nu}}\quad,\quad
 {\pi^\nu_a}^\ddag=\frac{\partial^r\mathcal{L}}{\partial\dot{\psi}^*_{a,\nu}}\
 ,\label{eqn2.1}
\end{eqnarray}
where $r$ means that the differentiation is performed from right
to left. We use the $\ddag$-notation to distinguish the canonical
momentum coming from the complex conjugate field from the one
coming form the original field, since they need not (and in fact
will not) be the same.

Using this prescription \eqref{eqn2.1} we obtain the canonical
momenta from our Lagrangians \eqref{eqn1.1a}, \eqref{eqn1.2a} and
\eqref{eqn1.2b}
\begin{eqnarray}
 \begin{array}{ll}
 \pi^0_{1}=0\ ,  &  \pi^n_{1}=-\dot{A}^n+\partial^n A^0\ ,  \\
 & \\
 \pi^0_{3/2}=0\ ,  &  {\pi^0_{3/2}}^\ddag=0\ ,  \\
 \pi^n_{3/2}=\frac{i}{2}\,\psi^{\dagger}_{k}\sigma^{kn}\ , &   {\pi^n_{3/2}}^\ddag=\frac{i}{2}\,\sigma^{nk}\psi_{k}\ ,\\
 & \\
 \pi^{00}_{2}=-\frac{1}{2}\,\partial_nh^{n0}\ ,  &
 \pi^{0m}_{2}=-\partial_nh^{nm}+\frac{1}{2}\,\partial^mh^{00}  \\
 \pi^{nm}_{2}=\frac{1}{2}\,\dot{h}^{nm}-\frac{1}{2}\,g^{nm}\dot{h}^k_k+\frac{1}{2}\,g^{nm}\partial_kh^{k0}\ ,   &
              \phantom{\pi^{0m}_{2}=}+\frac{1}{2}\,\partial^mh^n_n\ ,  \\
 \end{array}\nonumber\\\label{eqn2.2}
\end{eqnarray}
from which the velocities can be deduced
\begin{eqnarray}
 \dot{A}^n & = & -\pi^n_1+\partial^n A^0\ ,\nonumber\\
 \dot{h}^{nm}&=&2\pi^{nm}_{2}-g^{nm}{\pi_{2}}^k_k+\frac{1}{2}\,g^{nm}\partial_kh^{k0}\
 ,\nonumber\\
 \dot{h}^k_k&=&-{\pi_{2}}^k_k+\frac{3}{2}\,\partial_kh^{k0}\ ,\label{eqn2.3}
\end{eqnarray}
and the primary constraint equations
\begin{eqnarray}
 \begin{array}{ll}
 \theta^0_1=\pi^0_1\ ,  &  \\
  & \\
 \theta^0_{3/2}=\pi_{3/2}^0\ ,  &   {\theta_{3/2}^0}^\ddag={\pi_{3/2}^0}^\ddag\ ,\\
 \theta_{3/2}^n=\pi_{3/2}^n-\frac{i}{2}\,\psi^{\dagger}_{k}\sigma^{kn}\ , &
       {\theta_{3/2}^n}^\ddag={\pi_{3/2}^n}^\ddag-\frac{i}{2}\,\sigma^{nk}\psi_{k}\ , \\
  & \\
 \theta^{00}_{2}=\pi^{00}_2+\frac{1}{2}\,\partial_nh^{n0}\ , &
 \theta^{0m}_2=\pi^{0m}_2+\partial_nh^{nm}-\frac{1}{2}\,\partial^mh^{00}-\frac{1}{2}\,\partial^mh^n_n\ . \\
 \end{array}\nonumber\\\label{eqn2.4}
\end{eqnarray}
They vanish in the weak sense, to which we will come back below.

If we want these constraints to remain zero we impose the time
derivative of these constraints to be zero. We find it most easily
to define the time derivative via the Poisson bracket (Pb)
$\dot{\theta}=\left\{\theta,H\right\}_P+\partial\theta/\partial t$
\footnote{In practice it will turn out that the constraints do not
explicitly depend on time $t$}. We, therefore, need the
Hamiltonians.

Dirac has shown \cite{Dirac} that the Hamiltonian obtained in the
usual way is a weak equation \footnote{In constructing the usual
Hamiltonian explicit use can be made of the constraints, since
these are also weak equations} and does not give the correct EoM.
This can be repaired by adding the primary constraints
\eqref{eqn2.4} to the Hamiltonian by means of Lagrange multipliers
in order to make it a so-called strong equation. What we get is
\begin{eqnarray}
 H_w
&=&
 \int d^3x\ \mathcal{H}_w(x)=\int d^3x\left(\sum_{i}\pi_i\dot{q}_i-\mathcal{L}\right)\ ,\nonumber\\
 \mathcal{H}_{1,S}
&=&
 -\frac{1}{2}\,\pi_1^n\pi_{1,n}+\pi_1^n\partial_n A_0
 +\frac{1}{2}\,\partial_m A_n\partial^m A^n
 -\frac{1}{2}\,\partial_m A_n\partial^n A^m
 -\frac{1}{2}\,M_1^2A^0A_0
 \nonumber\\
&&
 -\frac{1}{2}\,M_1^2A^n A_n+\lambda_{1,0}\theta_1^0\ ,\nonumber\\
 \mathcal{H}_{3/2,S}
&=&
 \frac{1}{2}\,\epsilon^{\mu\nu\rho k}\bar{\psi}_{\mu}
 \gamma_5\gamma_{\rho}\left(\partial_{k}\psi_{\nu}\right)
 -\frac{1}{2}\,\epsilon^{\mu\nu\rho k}
 \left(\partial_{k}\bar{\psi}_{\mu}\right)\gamma_5\gamma_{\rho}\psi_{\nu}
 +M_{3/2}\bar{\psi}_{\mu}\sigma^{\mu\nu}\psi_{\nu}\nonumber\\
&&
 +\lambda_{3/2,0}\theta_{3/2}^0+\lambda_{3/2,n}\theta_{3/2}^{n}
 +\lambda_{3/2,0}^\ddag{\theta_{3/2}^0}^\ddag+\lambda_{3/2,n}^\ddag{\theta_{3/2}^{n}}^\ddag\ ,\nonumber\\
 \mathcal{H}_{2,S}
&=&
 \pi^{nm}_2\pi_{2,nm}-\frac{1}{2}\,{\pi_{2}}^n_n{\pi_{2}}^m_m+\frac{1}{2}\,{\pi_{2}}^n_n\partial^mh_{m0}
 -\frac{1}{2}\,\partial^kh^{n0}\partial_kh_{n0}-\frac{1}{4}\,\partial^kh^{nm}\partial_kh_{nm}
 \nonumber\\
&&
 +\frac{1}{8}\,\partial_nh^{n0}\partial^mh_{m0}+\frac{1}{2}\,\partial_nh^{nm}\partial^kh_{km}
 +\frac{1}{2}\,\partial_mh^{00}\partial^mh^n_n+\frac{1}{4}\,\partial_mh^n_n\partial^mh^k_k
 \nonumber\\
&&
 -\frac{1}{2}\,\partial_nh^{nm}\partial_mh_{00}
 -\frac{1}{2}\,\partial_nh^{nm}\partial_mh^k_k
 +\frac{1}{2}\,M_2^2h^{n0}h_{n0}+\frac{1}{4}\,M_2^2h^{nm}h_{nm}
 \nonumber\\
&&
 -\frac{1}{2}\,M_2^2h^{00}h^m_m
 -\frac{1}{4}\,M_2^2h^n_nh^m_m+\lambda_{2,00}\theta_2^{00}
 +\lambda_{2,0m}\theta_2^{0m}\ .\label{eqn2.5}
\end{eqnarray}
For the definition of the Pb we rely on \cite{senjan} and
\cite{Hiida}. There, it is defined as
\begin{eqnarray}
 \left\{E(x),F(y)\right\}_{P}
&=&
 \left[\frac{\partial^rE(x)}{\partial q_a(x)}\,\frac{\partial^lF(y)}{\partial p^a(y)}
 -(-1)^{n_En_F}\frac{\partial^rF(y)}{\partial q_a(y)}\,\frac{\partial^lE(x)}{\partial
 p^a(x)}\right]\delta^3(x-y)\ ,\label{eqn2.6}
\end{eqnarray}
where $n_E,n_F$ is 0 (1) in case $E(x),F(x)$ is even (odd). With
this form of the Pb \eqref{eqn2.6} we already anticipate that
bosons satisfy commutation relations and fermions anti-commutation
relations in a quantum theory.

Now, we can impose the time derivatives of the constraints
\eqref{eqn2.4} to be zero using (\ref{eqn2.5}) and (\ref{eqn2.6})
\begin{subequations}
\begin{eqnarray}
 \left\{\theta^0_1(x),H_{1,S}\right\}_P
&=&
 \partial_n\pi_1^n+M_1^2A^0=0\equiv\Phi^{0}_1(x)\ ,\\
&&
 \nonumber\\
 \left\{\theta_{3/2}^0(x),H_{3/2,S}\right\}_P
&=&
 \epsilon^{\mu 0\rho k}\left(\partial_{k}\bar{\psi}_{\mu}\right)
 \gamma_5\gamma_{\rho}-M_{3/2}\bar{\psi}_{\mu}\sigma^{\mu0}=0
 \equiv -{\Phi_{3/2}^{0}}^\ddag(x)\ ,\\
 \left\{{\theta_{3/2}^0}^\ddag(x),H_{3/2,S}\right\}_P
&=&
 -\epsilon^{\mu 0\rho k}\gamma^0\gamma_5\gamma_{\rho}
 \left(\partial_{k}\psi_{\mu}\right)+M_{3/2}\gamma^0\sigma^{0\mu}\psi_{\mu}=0
 \equiv -\Phi_{3/2}^{0}(x)\ ,\\
 \left\{\theta_{3/2}^n(x),H_{3/2,S}\right\}_P
&=&
 \epsilon^{\mu n\rho k}\left(\partial_{k}\bar{\psi}_{\mu}\right)\gamma_5\gamma_{\rho}
 -M_{3/2}\bar{\psi}_{\mu}\sigma^{\mu n}
 +i\lambda_{3/2,k}^\ddag\sigma^{kn}=0\ ,\label{eqn2.7a}\\*
 \left\{{\theta_{3/2}^n}^\ddag(x),H_{3/2,S}\right\}_P
&=&
 -\epsilon^{\mu n\rho k}\gamma^0\gamma_5\gamma_{\rho}\left(\partial_{k}\psi_{\mu}\right)
 +M_{3/2}\gamma^0\sigma^{n\mu}\psi_{\mu}
 +i\sigma^{nk}\lambda_{3/2,k}=0\ ,\label{eqn2.7b}\\
&&
 \nonumber\\
 \left\{\theta^{00}_2(x),H_{2,S}\right\}_P
&=&
 \frac{1}{2}\left[\left(\partial^k\partial_k+M_2^2\right)h^m_m
 -\partial_n\partial_mh^{nm}\right]=0
 \equiv \frac{1}{2}\,\Phi^{0}_2(x)\ ,\\
 \left\{\theta^{0m}_2(x),H_{2,Tot}\right\}_P
&=&
 2\partial_k\pi_2^{km}-\left(\partial^k\partial_k+M_2^2\right)h^{0m}=0
 \equiv \Phi^m_2(x)\ .
\end{eqnarray}\label{eqn2.7}
\end{subequations}
\footnote{If $\Phi$ is a constraint, then so is $a\Phi$. The
constants in front of the constraints in \eqref{eqn2.7} are chosen
for convenience and have no physical meaning.} In two cases
(\eqref{eqn2.7a} and \eqref{eqn2.7b}) Lagrange multipliers are
determined. In all other cases new, secondary, constraints are
obtained. We also impose the time derivatives of these secondary
constraints to be zero
\begin{subequations}
\begin{eqnarray}
 \left\{\Phi^0_1(x),H_{1,S}\right\}_P
&=&
 M_1^2(\partial_nA^n+\lambda_1^0)=0\ ,\label{eqn2.8a}\\
&&
 \nonumber\\
 \left\{\Phi_{3/2}^{0}(x),H_{3/2,S}\right\}_P
&=&
 \sigma^{nk}i\partial_n\lambda_{3/2,k}+M_{3/2}\gamma^k\lambda_{3/2,k}=0\ ,\label{eqn2.8b}\\
 \left\{{\Phi_{3/2}^{0}}^\ddag(x),H_{3/2,S}\right\}_P
&=&
 i\partial_k\lambda^\ddag_{3/2,n}\sigma^{nk}+M_{3/2}\lambda^\ddag_{3/2,k}\gamma^k=0\ ,\label{eqn2.8c}\\
&&
 \nonumber\\
 \left\{\Phi_2^{0}(x),H_{2,S}\right\}_P
&=&
 -2\partial_n\partial_m\pi_2^{nm}-M^2{\pi_2}^n_n+\left(\partial^k\partial_k
 +\frac{3}{2}\,M_{2}^2\right)\partial^nh_{n0}=0\nonumber\\
&\equiv&
 -\Phi_2^{(1)}(x)\ ,\label{eqn2.8d}\\
 \left\{\Phi^{m}_2(x),H_{2,S}\right\}_P
&=&
 -M_{2}^2\left[\lambda_2^{0m}+\partial_kh^{km}-\partial^mh^{00}-\partial^mh^n_n\right]
 =0\ .\label{eqn2.8e}
\end{eqnarray}
\end{subequations}
The first line \eqref{eqn2.8a} determines the Lagrange multiplier
$\lambda^0_1$. Since this was the only Lagrange multiplier in the
spin-1 case all Lagrange multipliers of this case are determined
and therefore all constraints are second class.

Equation \eqref{eqn2.8e} determines the Lagrange multiplier
$\lambda^{0m}_2$ and equation \eqref{eqn2.8d} brings about yet
another (tertiary) constraint. Its vanishing time derivative
yields
\begin{eqnarray}
 \left\{\Phi_2^{(1)}(x),H_{2,S}\right\}_P
&=&
 M_2^2\left[\left(2\partial^k\partial_k+\frac{3}{2}\,M_2^2\right)h^{00}
 +\left(\frac{3}{2}\,\partial^k\partial_k+M_2^2\right)h^n_n
 \right.\nonumber\\
&&
 \left.\phantom{M_2^2[}
 -\frac{3}{2}\,\partial_n\partial_mh^{nm}-2\partial_n\lambda^{n0}_2\right]=0\
 .\label{eqn2.9}
\end{eqnarray}
We see that we have in the spin-3/2 case as well as in the spin-2
case two equations involving the same Lagrange multipliers. In the
spin-3/2 case these are \eqref{eqn2.7b} and \eqref{eqn2.8b} for
$\lambda_{3/2,k}$ and \eqref{eqn2.7a} and \eqref{eqn2.8c} for
$\lambda^\ddag_{3/2,k}$. In the spin-2 case these are
\eqref{eqn2.8e} and \eqref{eqn2.9} for $\lambda_2^{n0}$. Combining
these equations for consistency, and using $\Phi_{3/2}^0$,
${\Phi_{3/2}^0}^\ddag$ as well as $\Phi_2^{0}$ as weakly vanishing
constraints, yields the last constraints
\begin{subequations}
\begin{eqnarray}
 \Phi_{3/2}^{(1)}
&=&
 \gamma^0\psi_0+\gamma^k\psi_k\ ,\label{eqn2.10a}\\
 {\Phi_{3/2}^{(1)}}^\ddag
&=&
 -\psi_0^\dagger\gamma^0+\psi_k^\dagger\gamma^k\
 ,\label{eqn2.10b}\\
&&
 \nonumber\\
 \Phi_2^{(2)}
&=&
 h^0_0+h^n_n\ ,\label{eqn2.10c}
\end{eqnarray}
\end{subequations}
It is important to note that these constraints are only obtained
when combining other results, as describes above. This is not done
in \cite{baaklini1}. Therefore these authors do not find
$\Phi_2^{(2)}$, leaving $\theta_2^{00}$ as a first class
constraint. Imposing vanishing time derivatives of these
constraints (\eqref{eqn2.10a}-\eqref{eqn2.10c})
\begin{eqnarray}
 \left\{\Phi_{3/2}^{(1)}(x),H_{3/2,S}\right\}_P
&=&
 -\gamma^0\lambda_{3/2,0}-\gamma^k\lambda_{3/2,k}=0\
 ,\nonumber\\
 \left\{{\Phi_{3/2}^{(1)}}^\ddag(x),H_{3/2,S}\right\}_P
&=&
 \lambda^\ddag_{3/2,0}\gamma^0-\lambda^\ddag_{3/2,k}\gamma^k=0\
 ,\nonumber\\
&&
 \nonumber\\*
 \left\{\Phi_2^{(2)}(x),H_{2,S}\right\}_P
&=&
 \lambda^{00}_2-{\pi_2}^k_k+\frac{3}{2}\,\partial_kh^{k0}=0\
 ,\label{eqn2.11}
\end{eqnarray}
determines the last Lagrange multipliers $\lambda_{3/2,0}$,
$\lambda^\ddag_{3/2,0}$ and $\lambda^{00}_2$.

In the massless spin-1 case the vanishing of the time-derivative
of $\Phi^0_1(x)$ would automatically be satisfied as can be seen
from (\ref{eqn2.8a}). In this case $\lambda_1^0$ would not be
determined which means that both constraints are first class.

We notice that in combining the equations that involve
$\lambda_{3/2,k}$ (\eqref{eqn2.7b}, \eqref{eqn2.8b}) and
$\lambda_{3/2,k}^\ddag$ (\eqref{eqn2.7a}, \eqref{eqn2.8c}) we
obtain the constraints $\Phi_{3/2}^{(1)}$ and
${\Phi_{3/2}^{(1)}}^\ddag$ being proportional to $M^2_{3/2}$. This
means that in the massless case these equations are already
consistent with each other and that $\lambda_{3/2,0}$ and
$\lambda_{3/2,0}^\ddag$ can not be determined leaving
$\theta^{0}_{3/2}$ and ${\theta^{0}_{3/2}}^\ddag$ to be a first
class constraint (\cite{senjan})\footnote{In this case also
$\partial_n{\theta_{3/2}^n}$ and
$\partial_n{\theta_{3/2}^n}^\ddag$ become first class.}.

The situation in the massless spin-2 case is even more clear. From
\eqref{eqn2.8e} and \eqref{eqn2.9} it is evident that the time
derivatives of $\Phi^{m}_2$ and $\Phi_2^{(1)}$ will already be
zero and that $\lambda^{0k}_2$ can not be determined. Therefore
$\Phi_2^{(2)}$ will not be obtained from which $\lambda^{00}_2$
also can not be determined, leaving $\theta^{00}_2$ and
$\theta^{0n}_2$ to be first class constraints (\cite{ghal,green})
\footnote{Actually all constraints become first class.}.

The fact that there are first class constraints (or undetermined
Lagrange multipliers) in the massless cases is a reflection of the
gauge symmetry. In the spin-1 and the spin-3/2 case only one
Lagrange multiplier is undetermined meaning there's only one gauge
symmetry (of course the massless spin-3/2 action is also invariant
under the hermitian gauge transformation, that's why
$\lambda_{3/2,k}^\ddag$ is also undetermined). In the massless
spin-2 case, however, there are two Lagrange multipliers
undetermined, meaning that there are two gauge symmetries as we
have mentioned before.\\

\noindent In the massive cases all Lagrange multipliers can be
determined, which means that all constraints are second class.
Therefore every constraint has at least one non-vanishing Pb with
another constraint. The complete set of constraints (primary,
secondary, \ldots) is
\begin{eqnarray}
\begin{array}{ll}
 \theta^0_1=\pi^0_1\ , & \Phi^0_1=\partial_n\pi^n_1+M_1^2 A^0\ , \\
  & \\
 \theta^0_{3/2}=\pi^0_{3/2}\ ,
 & {\theta^0_{3/2}}^\ddag={\pi^0}^\ddag\ , \\
 \Phi_{3/2}^{(1)}=\gamma\cdot\psi\ ,
 & {\Phi_{3/2}^{(1)}}^\ddag=-\psi_0^\dagger\gamma^0+\psi_k^\dagger\gamma^k\ ,\\
 \theta^n_{3/2}=\pi^n_{3/2}-\frac{i}{2}\,\psi^{\dagger}_k\sigma^{kn}\ ,
 & {\theta^n_{3/2}}^\ddag={\pi^n}^\ddag-\frac{i}{2}\,\sigma^{nk}\psi_k\ ,\\
 \Phi_{3/2}^{0}=-i\partial_k\sigma^{kl}\psi_l-M_{3/2}\gamma^k\psi_k\ ,
 & {\Phi_{3/2}^{0}}^\ddag=-\psi^\dagger_n\sigma^{nk}i\overleftarrow{\partial_k}-M_{3/2}\psi_k^\dagger\gamma^k\ ,\\
  & \\
 \theta^{00}_2=\pi^{00}_2+\frac{1}{2}\,\partial_nh^{n0}\ ,
 & \Phi^0_2=\left(\partial^k\partial_k+M_2^2\right)h^m_m-\partial_n\partial_mh^{nm}\ , \\
 \theta^{0m}_2=\pi^{0m}_2+\partial_nh^{nm}-\frac{1}{2}\,\partial^mh^{00}
 &  \Phi^m_2=2\partial_k\pi^{km}-(\partial^k\partial_k+M_2^2)h^{0m}\ ,\\
 \phantom{\theta^{0m}_2=}-\frac{1}{2}\,\partial^mh^n_n\ ,
 & \Phi^{(2)}_2=h^0_0+h^n_n\ ,\\
 \Phi^{(1)}_2=2\partial_n\partial_m\pi^{nm}_2+M_2^2{\pi_2}^n_n
 & \\
 \phantom{\Phi^m_2=}-\left(\partial^k\partial_k+\frac{3}{2}\,M_2^2\right)\partial^nh_{n0}\ , &
\end{array}\nonumber\\\label{eqn2.12}
\end{eqnarray}
We want to make linear combinations of constraints in order to
reduce the number of non-vanishing Pb among these constraints. In
the end we will arrive at a situation where every constraint has
only one non-vanishing Pb with another constraint. Therefore, we
make the following linear combinations
\begin{eqnarray}
 \tilde{\theta}^n_{3/2}
&=&
 \theta^n_{3/2}-\theta^0_{3/2}\gamma_0\gamma^n\ ,\nonumber\\
 \tilde{\Phi}^0_{3/2}
&=&
 \Phi^0_{3/2}+\left(-\partial_m+\frac{i}{2}\,M_{3/2}\gamma_m\right)\tilde{\theta}^m_{3/2}\ ,\nonumber\\
 \tilde{\theta}^{n\ddag}
&=&
 {\theta^n_{3/2}}^\ddag+\gamma^n\gamma^0{\theta_{3/2}^0}^\ddag\ ,\nonumber\\
 \tilde{\Phi}_{3/2}^{0\ddag}
&=&
 {\Phi_{3/2}^0}^\ddag+\tilde{\theta}_{3/2}^{m\ddag}\left(-\overleftarrow{\partial}_m+\frac{i}{2}\,M_{3/2}\gamma_m\right)
 \ ,\nonumber
\end{eqnarray}
\begin{eqnarray}
 \tilde{\Phi}^n_2
&=&
 \Phi_2^n-2\partial^n\theta_2^{00}\ ,\nonumber\\
 \tilde{\Phi}_2^0
&=&
 \Phi_2^0+2\partial_n\theta_2^{n0}\ ,\nonumber\\
 \tilde{\Phi}_2^{(1)}
&=&
 \Phi_2^{(1)}-(2\partial^k\partial_k+3M_2^2)\theta_2^{00}-2\partial_n\tilde{\Phi}_2^n\ .\label{eqn2.13}
\end{eqnarray}
The remaining non-vanishing Pb's are
\begin{eqnarray}
 \left\{\theta_1^0(x),\Phi^0_1(y)\right\}_P
&=&
 -M_1^2\delta^3(x-y)\ ,\nonumber\\
&&
 \nonumber\\
 \left\{\tilde{\theta}^n_{3/2}(x),\tilde{\theta}_{3/2}^{m\ddag}(y)\right\}_P
&=&
 -i\sigma^{mn}\delta^3(x-y)\ ,\nonumber\\
 \left\{\tilde{\Phi}_{3/2}^0(x),\tilde{\Phi}_{3/2}^{0\ddag}(y)\right\}_P
&=&
 -\frac{3i}{2}\,M_{3/2}^2\delta^3(x-y)\ ,\nonumber\\
 \left\{\theta^0_{3/2}(x),{\Phi^{(1)}_{3/2}}^{\ddag}(y)\right\}_P
&=&
 \gamma^0\delta^3(x-y)\ ,\nonumber\\
&&
 \nonumber\\
 \left\{\theta_2^{00}(x),\Phi^{(2)}_2(y)\right\}_P
&=&
 -\delta^3(x-y)\ ,\nonumber\\
 \left\{\tilde{\Phi}_2^{0}(x),\tilde{\Phi}_2^{(1)}(y)\right\}_P
&=&
 3M_2^4\,\delta^3(x-y)\ ,\nonumber\\
 \left\{\theta^{0n}_2(x),\tilde{\Phi}^m_2(y)\right\}_P
&=&
 M_2^2g^{nm}\,\delta^3(x-y)\ .\label{eqn2.14}
\end{eqnarray}
In a proper (quantum) theory we want the constraint to vanish.
Although, here, they vanish in the weak sense there still exist
non-vanishing Pb relations among them. This means in a quantum
theory that ETC and ETAC relations exist among the constraints.
We, therefore, introduce the new Pb \`a la Dirac \cite{Dirac}: The
Dirac bracket (Db), such that the Db among the constraints
vanishes
\begin{eqnarray}
 \left\{E(x),F(y)\right\}_{D}
&=&
 \left\{E(x),F(y)\right\}_{P}
 -\int d^3z_zd^3z_2\left\{E(x),\theta_a(z_1)\right\}_{P}
 \nonumber\\
&&
 \times
 C_{ab}(z_1-z_2)\left\{\theta_b(z_2),F(y)\right\}_{P}\
 ,\label{eqn2.15}
\end{eqnarray}
where the inverse functions $C_{ab}(z_1-z_2)$ are defined as
follows
\begin{eqnarray}
 \int d^3z\left\{\theta_a(x),\theta_c(z)\right\}_{P}C_{cb}(z-y)
 =\delta_{ab}\delta^3(x-y)\ ,\label{eqn2.16}
\end{eqnarray}
and can be deduced from \eqref{eqn2.14}.

The ETC and ETAC relations are obtained by multiplying the Db by a
factor of $i$ \footnote{Of course, this is not the only step to be
made when passing to a quantum theory. Also the fields should be
regarded as state operators, etc.}. What we get is
\begin{eqnarray}
 \left[A^0(x),A^n(y)\right]_0
&=&
 \frac{i\partial^n}{M_1^2}\,\delta^3(x-y)\ ,\nonumber\\
 \left[\dot{A}^0(x),A^0(y)\right]_0
&=&
 -\frac{i}{M_1^2}\,\partial^n\partial_n\,\delta^3(x-y)\
 ,\nonumber\\
 \left[\dot{A}^n(x),A^m(y)\right]_0
&=&
 i\left(g^{nm}+\frac{\partial^n\partial^m}{M_1^2}\right)\delta^3(x-y)\
 ,\nonumber
\end{eqnarray}
\begin{eqnarray}
 \left\{\psi^0(x),{\psi^0}^\dagger(y)\right\}_{0}
&=&
 -\frac{2}{3M_{3/2}^2}\,\nabla^2\,\delta^3(x-y)\ ,\nonumber\\
 \left\{\psi^0(x),{\psi^m}^\dagger(y)\right\}_{0}
&=&
 \frac{1}{M_{3/2}}\left[\frac{2}{3M_{3/2}}\left(i\gamma^k\partial_k\right)\gamma^0i\partial^m
 +\frac{1}{3}\left(i\gamma^k\partial_k\right)\gamma^0\gamma^m
 +\gamma^0i\partial^m\right]\delta^3(x-y)\ ,\nonumber\\
 \left\{\psi^n(x),{\psi^0}^\dagger(y)\right\}_{0}
&=&
 \frac{1}{M_{3/2}}\left[\frac{2}{3M_{3/2}}\left(i\gamma^k\partial_k\right)i\partial^n\gamma^0
 +\frac{1}{3}\gamma^n\gamma^0\left(i\gamma^k\partial_k\right)+i\partial^n\gamma^0\right]
 \delta^3(x-y)\ ,\nonumber\\
 \left\{\psi^n(x),{\psi^m}^\dagger(y)\right\}_{0}
&=&
 -\left[g^{nm}-\frac{1}{3}\,\gamma^{n}\gamma^{m}
 +\frac{2}{3M_{3/2}^2}\,\partial^n\partial^m
 +\frac{1}{3M_{3/2}}\left(\gamma^ni\partial^m
 \vphantom{\frac{A}{A}}
 -i\partial^n\gamma^m\right)\right]\delta^3(x-y)\ ,\nonumber
\end{eqnarray}
\begin{eqnarray}
 \left[h^{00}(x),h^{0l}(y)\right]_{0}
&=&
 \frac{4i}{3M_2^4}\,\partial^j\partial_j\partial^l\delta^3(x-y)\
 ,\nonumber\\
 \left[h^{0m}(x),h^{kl}(y)\right]_{0}
&=&
 \frac{-i}{M_2^2}\left[\frac{4}{3M^2}\,\partial^m\partial^k\partial^l
 -\frac{2}{3}\,\partial^mg^{kl}+\partial^kg^{ml}+\partial^lg^{mk}\right]
 \delta^3(x-y)\ ,\nonumber\\
 \left[\dot{h}^{00}(x),h^{00}(y)\right]_{0}
&=&
 -\frac{4i}{3M_2^4}\,\partial^i\partial_i\partial^j\partial_j\delta^3(x-y)\
 ,\nonumber\\
 \left[\dot{h}^{0m}(x),h^{0l}(y)\right]_{0}
&=&
 \frac{i}{M_2^2}\left[\frac{4}{3M_2^2}\,\partial^m\partial^l\,\partial^j\partial_j
 +\frac{1}{3}\,\partial^m\partial^l+\partial^j\partial_jg^{ml}\right]\delta^3(x-y)\
 ,\nonumber\\
 \left[\dot{h}^{00}(x),h^{kl}(y)\right]_{0}
&=&
 \frac{i}{M_2^2}\left[\frac{4}{3M_2^2}\,\partial^k\partial^l\,\partial^j\partial_j
 +2\partial^k\partial^l-\frac{2}{3}\,\partial^j\partial_jg^{kl}\right]\delta^3(x-y)\
 ,\nonumber\\
 \left[\dot{h}^{nm}(x),h^{kl}(y)\right]_{0}
&=&
 i\left[-g^{nk}g^{ml}-g^{nl}g^{mk}+\frac{2}{3}\,g^{nm}g^{kl}\right.\nonumber\\
&&
 \phantom{i[}-\frac{1}{M_2^2}\left(\partial^{n}\partial^{k}g^{ml}+\partial^{m}\partial^{k}g^{nl}
 +\partial^{n}\partial^{l}g^{mk}+\partial^{m}\partial^{l}g^{nk}\right)\nonumber\\
&&
 \left.\phantom{i[}+\frac{2}{3M_2^2}\left(\partial^{n}\partial^{m}g^{kl}+g^{nm}\partial^{k}\partial^{l}\right)
 -\frac{4}{3M_2^2}\,\partial^{n}\partial^{m}\partial^{k}\partial^{l}\right]
 \delta^3(x-y)\ .\label{eqn2.17}
\end{eqnarray}
This concludes the quantization of free, massive higher spin
($j=1,\,3/2,\,2$) fields. As a final remark we notice that the
ET(A)C relations in \eqref{eqn2.17} amongst the various components
of the spin-3/2, spin-2 field and their velocities are independent
of the choice of the parameter $A$ in \eqref{eqn1.1}.

\subsection{Propagators}\label{prop1}

Having quantized the free fields in the previous subsection
(section \ref{quant1}) we now want to obtain the propagators. In
order to do so we need to calculate the commutation relations for
non-equal times, which is done using the following identities as
solutions to the field equations (first column of (\ref{eqn1.3}))
\begin{eqnarray}
 A^{\mu}(x)
&=&
 \int d^3z\left[\partial^z_0\Delta(x-z;M_1^2)A^{\mu}(z)
 -\Delta(x-z;M_1^2)\partial^z_0A^{\mu}(z)\right]\ ,\nonumber\\
 \psi^{\mu}(x)
&=&
 i\int d^3z (i\slpart_x+M_{3/2})\gamma_0\Delta(x-z;M_{3/2}^2)\psi^{\mu}(z)\
 ,\nonumber\\
 h^{\mu\nu}(x)
&=&
 \int d^3z\left[\partial^z_0\Delta(x-z;M_2^2)h^{\mu\nu}(z)
 -\Delta(x-z;M_2^2)\partial^z_0h^{\mu\nu}(z)\right]\ .\label{eqn3.1}
\end{eqnarray}
Using these equations (\ref{eqn3.1}) and the ETC and ETAC
relations we obtained before (\ref{eqn2.17}) we calculate the
commutation relations for unequal times
\begin{eqnarray}
 \left[A^{\mu}(x),A^{\nu}(y)\right]
&=&
 -i\left(g^{\mu\nu}+\frac{\partial^\mu\partial^\nu}{M_1^2}\right)\Delta(x-y;M_1^2)
 = P^{\mu\nu}_1(\partial)\,i\Delta(x-y;M_1^2)\ ,\nonumber\\
 \left\{\psi^{\mu}(x),\bar{\psi}^{\nu}(y)\right\}
&=&
 -i\left(i\slpart+M_{3/2}\right)
 \left[g^{\mu\nu}-\frac{1}{3}\,\gamma^\mu\gamma^\nu+\frac{2\partial^\mu\partial^\nu}{3M_{3/2}^2}
 -\frac{1}{3M_{3/2}}\left(\gamma^\mu i\partial^\nu-\gamma^\nu
 i\partial^\mu\right)\right]\nonumber\\
&&
 \phantom{-i}\times\Delta(x-y;M_{3/2}^2)=
 \left(i\slpart+M_{3/2}\right)P_{3/2}^{\mu\nu}(\partial)\,i\Delta(x-y;M_{3/2}^2)
 \ ,\nonumber\\
 \left[h^{\mu\nu}(x),h^{\alpha\beta}(y)\right]
&=&
 i\left[g^{\mu\alpha}g^{\nu\beta}+g^{\mu\beta}g^{\nu\alpha}-\frac{2}{3}\,g^{\mu\nu}g^{\alpha\beta}\right.\nonumber\\
&&
 \phantom{i[}+\frac{1}{M_2^2}\left(\partial^{\mu}\partial^{\alpha}g^{\nu\beta}+\partial^{\nu}\partial^{\alpha}g^{\mu\beta}
 +\partial^{\mu}\partial^{\beta}g^{\nu\alpha}+\partial^{\nu}\partial^{\beta}g^{\mu\alpha}\right)\nonumber\\
&&
 \left.\phantom{i[}-\frac{2}{3M_2^2}\left(\partial^{\mu}\partial^{\nu}g^{\alpha\beta}+g^{\mu\nu}\partial^{\alpha}\partial^{\beta}\right)
 +\frac{4}{3M_2^2}\,\partial^{\mu}\partial^{\nu}\partial^{\alpha}\partial^{\beta}\right]\Delta(x-y;M_2^2)\nonumber\\
&=&
 2P^{\mu\nu\alpha\beta}_2(\partial)\,i\Delta(x-y;M_2^2)\ ,\label{eqn3.2}
\end{eqnarray}
where the $P_j(\partial),\ j=1,\,3/2,\,2$ are the (on mass shell)
spin projection operators. The factor $2$ in the last line of
(\ref{eqn3.2}) can be transformed away by redefining the spin-2
field. Equation (\ref{eqn3.2}) yields for the propagators
\begin{eqnarray}
 D_{F}^{\mu\nu}(x-y)
&=&
 -i<0|T\left[A^{\mu}(x)A^{\nu}(y)\right]|0>\nonumber\\
&=&
 -i\theta(x^0-y^0)P^{\mu\nu}_1(\partial)\Delta^{(+)}(x-y;M_1^2)
 -i\theta(y^0-x^0)P^{\mu\nu}_1(\partial)\Delta^{(-)}(x-y;M_1^2)\nonumber\\
&=&
 P^{\mu\nu}_1(\partial)\Delta_F(x-y;M_1^2)-i\delta^{\mu}_0\delta^{\nu}_0\,\delta^4(x-y)
 \ .\label{eqn3.3}\\\nonumber\\
 S_{F}^{\mu\nu}(x-y)
&=&
 -i<0|T\left(\psi^{\mu}(x)\bar{\psi}^{\nu}(y)\right)|0>\nonumber\\
&=&
 -i\theta(x^0-y^0)\left(i\slpart+M_{3/2}\right)P^{\mu\nu}_{3/2}(\partial)\Delta^{(+)}(x-y;M_{3/2}^2)\nonumber\\
&&
 -i\theta(y^0-x^0)\left(i\slpart+M_{3/2}\right)P^{\mu\nu}_{3/2}(\partial)\Delta^{(-)}(x-y;M_{3/2}^2)\nonumber\\
&=&
 \left(i\slpart+M_{3/2}\right)P^{\mu\nu}_{3/2}(\partial)
 \Delta_F(x-y;M_{3/2}^2)\nonumber\\
&&
 -\gamma_0\left[\frac{2}{3M_{3/2}^2}\left(\delta^{\mu}_0\delta^{\nu}_m+
 \delta^{\nu}_0\delta^{\mu}_m\right)i\partial^m+\frac{1}{3M_{3/2}}\left(
 \delta^{\mu}_m\delta^{\nu}_0-\delta^{\nu}_m\delta^{\mu}_0\right)\gamma^m\right]
 \delta^4(x-y)\nonumber\\
&&
 -\frac{2}{3M_{3/2}^2}\left(i\slpart+M_{3/2}\right)\delta^{\mu}_0\delta^{\nu}_0\delta^4(x-y)\ .\label{eqn3.4}
 \\\nonumber\\
 D_{F}^{\mu\nu\alpha\beta}(x-y)
&=&
 -i<0|T\left[h^{\mu\nu}(x)h^{\alpha\beta}(y)\right]|0>\nonumber\\
&=&
 -i\theta(x^0-y^0)2P^{\mu\nu\alpha\beta}_2(\partial)\Delta^{(+)}(x-y;M_2^2)
 -i\theta(y^0-x^0)2P^{\mu\nu\alpha\beta}_2(\partial)\Delta^{(-)}(x-y;M_2^2)\nonumber\\
&=&
 2P^{\mu\nu\alpha\beta}_2(\partial)\Delta_F(x-y;M_2^2)\nonumber\\
&&
 +\frac{1}{M_2^2}\left[\vphantom{\frac{A}{A}}
 \delta^{\mu}_0\delta^{\alpha}_0g^{\nu\beta}+\delta^{\nu}_0\delta^{\alpha}_0g^{\mu\beta}
 +\delta^{\mu}_0\delta^{\beta}_0g^{\nu\alpha}+\delta^{\nu}_0\delta^{\beta}_0g^{\mu\alpha}\right.\nonumber\\
&&
 \phantom{+\frac{1}{M_2^2}\left[\right.}
 -\frac{2}{3}\left(\delta^{\mu}_0\delta^{\nu}_0g^{\alpha\beta}+g^{\mu\nu}\delta^{\alpha}_0\delta^{\beta}_0\right)
 +\frac{4}{3}\left(
 \delta^{\mu}_0\delta^{\nu}_0\delta^{\alpha}_0\delta^{\beta}_0(\partial^0\partial_0-\partial^k\partial_k-M_2^2)
 \right.\nonumber\\
&&
 \phantom{+\frac{1}{M_2^2}\left[\right.}
 +\delta^{\mu}_0\delta^{\nu}_0\delta^{\alpha}_0\delta^{\beta}_b\partial^0\partial^b
 +\delta^{\mu}_0\delta^{\nu}_0\delta^{\alpha}_a\delta^{\beta}_0\partial^0\partial^a
 +\delta^{\mu}_0\delta^{\nu}_n\delta^{\alpha}_0\delta^{\beta}_0\partial^0\partial^n
 +\delta^{\mu}_m\delta^{\nu}_0\delta^{\alpha}_0\delta^{\beta}_0\partial^0\partial^m\nonumber\\
&&
 \phantom{+\frac{1}{M_2^2}\left[\right.}
 +\delta^{\mu}_0\delta^{\nu}_0\delta^{\alpha}_a\delta^{\beta}_b\partial^a\partial^b
 +\delta^{\mu}_0\delta^{\nu}_n\delta^{\alpha}_0\delta^{\beta}_b\partial^n\partial^b
 +\delta^{\mu}_m\delta^{\nu}_0\delta^{\alpha}_0\delta^{\beta}_b\partial^m\partial^b
 +\delta^{\mu}_0\delta^{\nu}_n\delta^{\alpha}_a\delta^{\beta}_0\partial^n\partial^a\nonumber\\
&&
 \phantom{+\frac{1}{M_2^2}\left[\right.}\left.\left.
 +\delta^{\mu}_m\delta^{\nu}_0\delta^{\alpha}_a\delta^{\beta}_0\partial^m\partial^a
 +\delta^{\mu}_m\delta^{\nu}_n\delta^{\alpha}_0\delta^{\beta}_0\partial^m\partial^n
 \right)\right]\delta^4(x-y)\ .\label{eqn3.5}
\end{eqnarray}
The use of $\Delta^{(+)}(x-y)$ and $\Delta^{(-)}(x-y)$ is similar
to what is written in \cite{Bjorken} in case of scalar fields
\begin{eqnarray}
 <0|\phi(x)\phi(y)|0>&=&\Delta^{(+)}(x-y)\ ,\nonumber\\
 <0|\phi(y)\phi(x)|0>&=&\Delta^{(-)}(x-y)\ .\label{eqn3.6}
\end{eqnarray}
As can be seen from (\eqref{eqn3.3}-\eqref{eqn3.5}) the
propagators are not covariant; they contain non-covariant, local
terms, as is mentioned in for instance \cite{weinberg}.

\section{Auxiliary Fields}\label{aux}

The goal of this section is to come to covariant propagators. The
way we do this is to introduce auxiliary fields. Since we also
allow for mass terms we have extra parameters which can be seen as
gauge parameters. We discuss certain choices of these parameters.
Also we discuss the massless limits of the propagators in section
\ref{massless} and give momentum representations of the fields in
section \ref{momrep}. Apart from that, the organization of this
section is exactly the same as the previous one (section
\ref{ffields}).

\subsection{Equations of Motion}\label{eom2}

As a starting point we take the Lagrangians \eqref{eqn1.1a},
\eqref{eqn1.2a} and \eqref{eqn1.2b}. To these Lagrangians we add
auxiliary fields coupled to the gauge conditions of the massless
theory, as discussed in the text below \eqref{eqn1.3}. We also
allow for mass terms of these auxiliary fields, which introduces
parameters to be seen as gauge parameters
\begin{subequations}
\begin{eqnarray}\label{eqn4.1}
 \mathcal{L}_{B}
&=&
 \mathcal{L}_{1}+M_1B\partial^{\mu}A_{\mu}+\frac{1}{2}\,aM_1^2 B^2\
 ,\label{eqn4.1a}\\
 \mathcal{L}_{\chi}
&=&
 \mathcal{L}_{3/2}+M_{3/2}\bar{\chi}\gamma^{\mu}\psi_{\mu}+M_{3/2}\bar{\psi}_{\mu}\gamma^{\mu}\chi
 +bM_{3/2}\bar{\chi}\chi\ ,\label{eqn4.1b}\\
 \mathcal{L}_{\eta\epsilon}
&=&
 \mathcal{L}_2+M_2\partial_\mu h^{\mu\nu}\eta_\nu+M_2^2h^{\mu}_{\mu}\epsilon
 +\frac{1}{2}\,cM_2^2\eta^\mu\eta_\mu\ .\label{eqn4.1c}
\end{eqnarray}
\end{subequations}
In \eqref{eqn4.1c} we did not allow for a mass term for the
$\epsilon$ field. We will come back to this point below.

These Lagrangians (\eqref{eqn4.1a}-\eqref{eqn4.1c}) lead to the
following EoM's.
\begin{eqnarray}
 \left(\Box+M_1^2\right)A^{\mu}
&=&
 (1-a)M_1\partial^{\mu}B\ ,\nonumber\\
 \left(\Box+M_B^2\right)\left(\Box+M_1^2\right)A^{\mu}
&=&
 0\ ,\nonumber\\*
 \left(\Box+M_B^2\right)B
&=&0\ ,\label{eqn4.2}
\end{eqnarray}
where $M_B^2=aM_1^2$. Furthermore we have the constraint relation
$\partial^{\mu}A_{\mu}=-aM_1B$.
\begin{eqnarray}
 \left(i\slpart-M_{3/2}\right)\psi_{\mu}
&=&
 -\frac{b+2}{2}\,M_{3/2}\gamma_\mu\chi-bi\partial_\mu\chi\ ,\nonumber\\
 \left(i\slpart+M_\chi\right)\left(i\slpart-M_{3/2}\right)\psi_{\mu}
&=&
 -(3b^2+5b+2)M_{3/2}i\partial_\mu\chi\ ,\nonumber\\
 \left(\Box+M_\chi^2\right)\left(i\slpart-M_{3/2}\right)\psi_{\mu}
&=&
 0\ ,\nonumber\\
 \left(i\slpart-M_{\chi}\right)\chi
&=&
 0\ ,\label{eqn4.3}
\end{eqnarray}
where $M_\chi=(3b/2+2)M_{3/2}$. The auxiliary field is related to
the original spin-3/2 field via the equations
$\gamma\cdot\psi=-b\chi$ and
$i\partial\cdot\psi=-\frac{1}{2}\,(1+b)(3b+4)M_{3/2}\chi$.
\begin{eqnarray}
 \left(\Box+M_2^2\right)h^{\mu\nu}
&=&
 -\left(1+c\right)M_2\left(\partial^\mu\eta^\nu+\partial^\nu\eta^\mu\right)
 +\frac{2\left(1+c\right)}{1-c}M_2^2g^{\mu\nu}\epsilon\ ,\nonumber\\
 \left(\Box+M_\eta^2\right)\left(\Box+M_2^2\right)h^{\mu\nu}
&=&
 \frac{2\left(1+c\right)^2}{1-c}\,M_2^2
 \left(2\partial^\mu\partial^\nu-\frac{c}{3+c}M_2^2g^{\mu\nu}\right)\epsilon\ ,\nonumber\\
 \left(\Box+M_\epsilon^2\right)\left(\Box+M_\eta^2\right)\left(\Box+M_2^2\right)h^{\mu\nu}
&=&
 0\ ,\nonumber\\
 \left(\Box+M_\eta^2\right)\eta^\mu
&=&
 -\frac{2\left(1+c\right)}{1-c}\,M_2\partial^\mu\epsilon\ ,\nonumber\\
 \left(\Box+M_\epsilon^2\right)\left(\Box+M_\eta^2\right)\eta^\mu
&=&
 0\ ,\nonumber\\
 \left(\Box+M_\epsilon^2\right)\epsilon
&=&
 0\ ,\label{eqn4.4}
\end{eqnarray}
where $M^2_\eta=-cM_2^2$ and $M_\epsilon^2=-\frac{2c}{3+c}M_2^2$.
The constraint relations are $h^\mu_\mu=0$, $\partial_\mu
h^{\mu\nu}=-cM_2\eta^\nu$ and
$\partial\cdot\eta=\frac{4M_2}{1-c}\,\epsilon$

From the last line of \eqref{eqn4.4} we see that the
$\epsilon$-field is a free Klein-Gordon field. This equation comes
about quite naturally from the Euler-Lagrange equations. This
would not be so if we allowed for a mass term of this
$\epsilon$-field in the Lagrangian \eqref{eqn4.1c}. Then it must
be imposed that $\epsilon$ is a free Klein-Gordon field which
makes the calculations unnatural and unnecessary difficult.

\subsection{Quantization}\label{quant2}

As mentioned before the quantization procedure runs exactly the
same as in the previous section (section \ref{quant1}). We,
therefore, determine the canonical momenta to be
\begin{eqnarray}
 \begin{array}{ll}
 \pi^0_{1}=M_1B\ ,  &  \pi_B=0\ ,  \\
 \pi^n_{1}=-\dot{A}^n+\partial^n A^0\ ,  &  \\
 & \\
 \pi^0_{3/2}=0\ ,  &  {\pi^0_{3/2}}^\ddag=0\ ,  \\
 \pi^n_{3/2}=\frac{i}{2}\,\psi^{\dagger}_{k}\sigma^{kn}\ , &   {\pi^n_{3/2}}^\ddag=\frac{i}{2}\,\sigma^{nk}\psi_{k}\ ,\\
 \pi_{\chi}=0\ ,  &  \pi_{\chi}^\ddag=0\ , \\
 & \\
 \pi^{00}_{2}=-\frac{1}{2}\,\partial_nh^{n0}+M_2\eta^0\ ,  &  \pi^{0}_{\eta}=0\ ,  \\
 \pi^{0m}_{2}=-\partial_nh^{nm}+\frac{1}{2}\,\partial^mh^{00}+\frac{1}{2}\,\partial^mh^n_n+M_2\eta^m\ ,   &  \pi^{m}_{\eta}=0\ ,  \\
 \pi^{nm}_{2}=\frac{1}{2}\,\dot{h}^{nm}-\frac{1}{2}\,g^{nm}\dot{h}^k_k+\frac{1}{2}\,g^{nm}\partial_kh^{k0}\ ,   &  \pi_{\epsilon}=0\ ,  \\
 \end{array}\nonumber\\\label{eqn5.1}
\end{eqnarray}
from which we deduce the velocities
\begin{eqnarray}
 \dot{A}^n & = & -\pi^n_1+\partial^n A^0\ ,\nonumber\\
 \dot{h}^{nm}&=&2\pi^{nm}_{2}-g^{nm}{\pi_{2}}^k_k+\frac{1}{2}\,g^{nm}\partial_kh^{k0}\
 ,\nonumber\\
 \dot{h}^k_k&=&-{\pi_{2}}^k_k+\frac{3}{2}\,\partial_kh^{k0}\ .\label{eqn5.2}
\end{eqnarray}
These velocities are the same as in the previous section (see
\eqref{eqn2.3}). The primary constraints are
\begin{eqnarray}
 \begin{array}{ll}
 \theta^0_1=\pi^0_1-M_1B\ , &  \theta_B=\pi_B\ ,  \\
  & \\
 \theta^0_{3/2}=\pi_{3/2}^0\ ,  &   {\theta_{3/2}^0}^\ddag={\pi_{3/2}^0}^\ddag\ ,\\
 \theta_{3/2}^n=\pi_{3/2}^n-\frac{i}{2}\,\psi^{\dagger}_{k}\sigma^{kn}\ , &
       {\theta_{3/2}^n}^\ddag={\pi_{3/2}^n}^\ddag-\frac{i}{2}\,\sigma^{nk}\psi_{k}\ , \\
 \theta_\chi=\pi_{\chi}\ ,  &  \theta_{\chi}^\ddag=\pi_{\chi}^\ddag\ , \\
  & \\
 \theta^{00}_{2}=\pi^{00}_2+\frac{1}{2}\,\partial_nh^{n0}-M_2\eta^0\ , & \theta^{0}_{\eta}=\pi^{0}_{\eta}\ ,\\
 \theta^{0m}_2=\pi^{0m}_2+\partial_nh^{nm}-\frac{1}{2}\,\partial^mh^{00}\qquad  & \theta^{m}_{\eta}=\pi^{m}_{\eta}\ ,\\
 \phantom{\theta^{0m}_2=}-\frac{1}{2}\,\partial^mh^n_n-M_2\eta^m\ , &  \theta_{\epsilon}=\pi_{\epsilon}\ . \\
 \end{array}\nonumber\\\label{eqn5.3}
\end{eqnarray}
Having determined the canonical momenta, the velocities and the
primary constraints we determine the (strong) Hamiltonians to be
\begin{eqnarray}
 \mathcal{H}_{B,S}
&=&
 -\frac{1}{2}\,\pi_1^n\pi_{1,n}+\pi_1^n\partial_n A_0
 +\frac{1}{2}\,\partial_m A_n\partial^m A^n
 -\frac{1}{2}\,\partial_m A_n\partial^n A^m
 -\frac{1}{2}\,M_1^2A^0A_0\nonumber\\
&&
 -\frac{1}{2}\,M_1^2A^n A_n-M_1B\partial^{m}A_{m}-\frac{1}{2}\,aM_1^2 B^2
 +\lambda_{1,0}\theta_1^0+\lambda_B\theta_B\ ,\nonumber\\
 \mathcal{H}_{\chi,S}
&=&
 \frac{1}{2}\,\epsilon^{\mu\nu\rho k}\bar{\psi}_{\mu}
 \gamma_5\gamma_{\rho}\left(\partial_{k}\psi_{\nu}\right)
 -\frac{1}{2}\,\epsilon^{\mu\nu\rho k}
 \left(\partial_{k}\bar{\psi}_{\mu}\right)\gamma_5\gamma_{\rho}\psi_{\nu}
 +M_{3/2}\bar{\psi}_{\mu}\sigma^{\mu\nu}\psi_{\nu}\nonumber\\
&&
 -M_{3/2}\bar{\chi}\gamma^{\mu}\psi_{\mu}-M_{3/2}\bar{\psi}_{\mu}\gamma^{\mu}\chi
 -bM_{3/2}\bar{\chi}\chi+\lambda_{3/2,0}\theta_{3/2}^0+\lambda_{3/2,n}\theta_{3/2}^{n}\nonumber\\
&&
 +\lambda_{3/2,0}^\ddag{\theta_{3/2}^0}^\ddag+\lambda_{3/2,n}^\ddag{\theta_{3/2}^{n}}^\ddag
 +\lambda_\chi\theta_\chi+\lambda_\chi^\ddag\theta_{\chi}^\ddag
 \ ,\nonumber\\
 \mathcal{H}_{\eta\epsilon,S}
&=&
 \pi^{nm}_2\pi_{2,nm}-\frac{1}{2}\,{\pi_{2}}^n_n{\pi_{2}}^m_m+\frac{1}{2}\,{\pi_{2}}^n_n\partial^mh_{m0}
 -\frac{1}{2}\,\partial^kh^{n0}\partial_kh_{n0}
 -\frac{1}{4}\,\partial^kh^{nm}\partial_kh_{nm}
 \nonumber\\
&&
 +\frac{1}{8}\,\partial_nh^{n0}\partial^mh_{m0}+\frac{1}{2}\,\partial_nh^{nm}\partial^kh_{km}
 +\frac{1}{2}\,\partial_mh^{00}\partial^mh^n_n+\frac{1}{4}\,\partial_mh^n_n\partial^mh^k_k
 \nonumber\\
&&
 -\frac{1}{2}\,\partial_nh^{nm}\partial_mh_{00}-\frac{1}{2}\,\partial_nh^{nm}\partial_mh^k_k
 +\frac{1}{2}\,M_2^2h^{n0}h_{n0}+\frac{1}{4}\,M_2^2h^{nm}h_{nm}
 \nonumber\\
&&
 -\frac{1}{2}\,M_2^2h^{00}h^m_m-\frac{1}{4}\,M_2^2h^n_nh^m_m
 -\frac{1}{2}\,cM_2^2\eta^\mu\eta_\mu -M_2\partial_n h^{n0}\eta_0
 -M_2\partial_n h^{nm}\eta_m\nonumber\\
&&
 -M_2^2h^{0}_{0}\epsilon-M_2^2h^{k}_{k}\epsilon
 +\lambda_{2,00}\theta_2^{00}
 +\lambda_{2,0m}\theta_2^{0m}+\lambda_{0,\eta}\theta^{0}_{\eta}
 +\lambda_{m,\eta}\theta^{m}_{\eta}+\lambda_{\epsilon}\theta_{\epsilon}
 \ .\label{eqn5.4}
\end{eqnarray}
With this Hamiltonians \eqref{eqn5.4} and with the definition of
the Pb in \eqref{eqn2.6} we impose the time-derivatives of the
constraints \eqref{eqn5.3} to be zero
\begin{subequations}
\begin{eqnarray}
 \left\{\theta^0_1(x),H_{B,S}\right\}_P
&=&
 \partial_n\pi_1^n+M_1^2A^0-M_1\lambda_B=0\ ,\label{eqn5.5a}\\
 \left\{\theta_B(x),H_{B,S}\right\}_P
&=&
 M_1\partial^{m}A_{m}+aM_1^2 B+M_1\lambda_{1,0}=0\ ,\label{eqn5.5b}
\end{eqnarray}\label{eqn5.5}
\end{subequations}
\begin{subequations}
\begin{eqnarray}
 \left\{\theta_{3/2}^0(x),H_{\chi,S}\right\}_P
&=&
 \epsilon^{\mu 0\rho k}\left(\partial_{k}\bar{\psi}_{\mu}\right)
 \gamma_5\gamma_{\rho}-M_{3/2}\bar{\psi}_{\mu}\sigma^{\mu0}+M_{3/2}\bar{\chi}\gamma^0
 =0\equiv -{\Phi_{3/2}^{0}}^\ddag(x)\ ,\\
 \left\{{\theta_{3/2}^0}^\ddag(x),H_{\chi,S}\right\}_P
&=&
 -\epsilon^{\mu 0\rho k}\gamma^0\gamma_5\gamma_{\rho}
 \left(\partial_{k}\psi_{\mu}\right)+M_{3/2}\gamma^0\sigma^{0\mu}\psi_{\mu}-M_{3/2}\chi=0
 \nonumber\\
&\equiv&
 -\Phi_{3/2}^{0}(x)\ ,\\
 \left\{\theta_{3/2}^n(x),H_{\chi,S}\right\}_P
&=&
 \epsilon^{\mu n\rho k}\left(\partial_{k}\bar{\psi}_{\mu}\right)\gamma_5\gamma_{\rho}
 -M_{3/2}\bar{\psi}_{\mu}\sigma^{\mu n}+M\bar{\chi}\gamma^n
 +i\lambda_{3/2,k}^\ddag\sigma^{kn}=0\ ,\label{eqn5.6c}\\
 \left\{{\theta_{3/2}^n}^\ddag(x),H_{\chi,S}\right\}_P
&=&
 -\epsilon^{\mu n\rho k}\gamma^0\gamma_5\gamma_{\rho}\left(\partial_{k}\psi_{\mu}\right)
 +M_{3/2}\gamma^0\sigma^{n\mu}\psi_{\mu}
 -M\gamma^0\gamma^n\chi+i\sigma^{nk}\lambda_{3/2,k}=0,\label{eqn5.6d}\\
 \left\{\theta_{\chi}(x),H_{\chi,S}\right\}_P
&=&
 M_{3/2}\bar{\psi}\cdot\gamma+bM_{3/2}\bar{\chi}=0\equiv-M_{3/2}\Phi_\chi^\ddag\gamma^0\ ,\\
 \left\{\theta_{\chi}^\ddag(x),H_{\chi,S}\right\}_P
&=&
 -M_{3/2}\gamma^0\gamma\cdot\psi-bM_{3/2}\gamma^0\chi=0
 \equiv -M_{3/2}\gamma^0\Phi_\chi\ ,
\end{eqnarray}\label{eqn5.6}
\end{subequations}
\begin{subequations}
\begin{eqnarray}
 \left\{\theta^{00}_2(x),H_{\eta\epsilon,S}\right\}_P
&=&
 -M_2\lambda_\eta^0+\frac{1}{2}\left(\partial^k\partial_k+M_2^2\right)h^m_m
 -\frac{1}{2}\partial_n\partial_mh^{nm}
 +M_2^2\epsilon=0\ ,\label{eqn5.7e}\\
 \left\{\theta^{0m}_2(x),H_{\eta\epsilon,S}\right\}_P
&=&
 2\partial_k\pi_2^{km}-\left(\partial^k\partial_k+M_2^2\right)h^{0m}
 -M_2\partial^m\eta^0
 -M_2\lambda_\eta^m=0\ ,\label{eqn5.7f}\\
 \left\{\theta^{0}_\eta(x),H_{\eta\epsilon,S}\right\}_P
&=&
 \partial_nh^{n0}+\lambda_2^{00}+cM_2\eta^0=0\ ,\label{eqn5.7g}\\
 \left\{\theta^{m}_\eta(x),H_{\eta\epsilon,S}\right\}_P
&=&
 \partial_nh^{nm}+\lambda_2^{0m}+cM_2\eta^m=0\ ,\label{eqn5.7h}\\
 \left\{\theta_\epsilon(x),H_{\eta\epsilon,S}\right\}_P
&=&
 M_2^2\left[h^0_0+h^n_n\right]=0\equiv M_2^2\Phi_\eta\ ,
\end{eqnarray}\label{eqn5.7}
\end{subequations}
Equations \eqref{eqn5.5a}, \eqref{eqn5.5b}, \eqref{eqn5.6c},
\eqref{eqn5.6d} and \eqref{eqn5.7e}-\eqref{eqn5.7h} determine the
Lagrange multipliers
$\lambda_B,\lambda_{1,0},\lambda_{3/2,k}^\ddag,$
$\lambda_{3/2,k},\lambda_\eta^0,
\lambda_\eta^m,\lambda_2^{00},\lambda_2^{0m}$, respectively. All
other equations in \eqref{eqn5.5}, \eqref{eqn5.6} and
\eqref{eqn5.7} yield new (secondary) constraints. Imposing their
time derivatives to be zero, yields
\begin{eqnarray}
 \left\{\Phi_{3/2}^{0}(x),H_{\chi,S}\right\}_P
&=&
 \sigma^{nk}i\partial_n\lambda_k+M_{3/2}\gamma^k\lambda_{3/2,k}
 -M_{3/2}\lambda_\chi=0\ ,\nonumber\\
 \left\{{\Phi_{3/2}^{0}}^\ddag(x),H_{\chi,S}\right\}_P
&=&
 i\partial_n\lambda_{3/2,k}^\ddag\sigma^{kn}+M_{3/2}\lambda_{3/2,k}^\ddag\gamma^k
 +M_{3/2}\lambda_\chi^\ddag=0\ ,\nonumber\\
 \left\{\Phi_{\chi}(x),H_{\chi,S}\right\}_P
&=&
 -b\lambda_\chi-\gamma^0\lambda_{3/2,0}-\gamma^n\lambda_{3/2,n}=0\ ,\nonumber\\
 \left\{\Phi_{\chi}^\ddag(x),H_{\chi,S}\right\}_P
&=&
 b\lambda_\chi^\ddag+\lambda_{3/2,0}^\ddag\gamma^0-\lambda_{3/2,n}^\ddag\gamma^n=0
 \ ,\label{eqn5.8}
\end{eqnarray}
\begin{eqnarray}
 \left\{\Phi_\eta(x),H_{\eta\epsilon}\right\}_P
&=&
 -{\pi_{2}}^{k}_k+\frac{1}{2}\,\partial_nh^{n0}-cM_{2}\eta^0=0
 =-\Phi^{(1)}_{2}\ .\label{eqn5.9}
 \qquad\label{eqn5.9}
\end{eqnarray}
The equations in \eqref{eqn5.8} determine the Lagrange multipliers
$\lambda_\chi$, $\lambda_\chi^\ddag$, $\lambda_{3/2,0}$ and
$\lambda_{3/2,0}^\ddag$. Equation \eqref{eqn5.9} yields yet
another (tertiary) constraint. Imposing its time derivative to be
zero
\begin{eqnarray}
\left\{\Phi^{(1)}_2(x),H_{\eta\epsilon}\right\}_P &=&
 \partial^k\partial_kh^{00}+\frac{1}{2}\,\partial^k\partial_kh^m_m
 -\frac{1}{2}\,\partial_n\partial_mh^{nm}+\frac{3}{2}\,M_2^2h^{00}+M_2^2h^m_m
 \nonumber\\
&&
 -M_2\partial^k\eta_k-\partial_m\lambda_{2}^{m0}+3M_2^2\epsilon
 +cM_2\lambda_\eta^0=0\ ,\label{eqn5.10}
\end{eqnarray}
gives an equation for $\lambda_\eta^0$. Since we already had an
equation determining $\lambda_\eta^0$ \eqref{eqn5.7e} we combine
both equations for consistency and use $\Phi_\eta$ as a weakly
vanishing constraint. What we get is the last constraint
\begin{eqnarray}
 \Phi_2^{(2)}
&=&
 -\partial_n\partial_mh^{nm}+\left(\partial^k\partial_k+M_2^2\right)h^m_m
 +2M_2\partial^k\eta_k
 -2\left(\frac{3+c}{1-c}\right)M_2^2\epsilon\ ,\nonumber\\
 \left\{\Phi_2^{(2)}(x),H_{\eta\epsilon,S}\right\}_P
&=&
 -2\partial_n\partial_m\pi_2^{nm}-M_2^2{\pi_2}^k_k
 +\left(\partial^k\partial_k+\frac{3}{2}\,M_2^2\right)\partial_nh^{n0}
 +2M_2\partial_k\lambda^k_\eta
 \nonumber\\
&&
 -2\left(\frac{3+c}{1-c}\right)M_2^2\lambda_\epsilon=0\ .\label{eqn5.11}
\end{eqnarray}
As can be seen in \eqref{eqn5.11} imposing the time derivative of
$\Phi_2^{(2)}$ to be zero determines the remaining Lagrange
multiplier $\lambda_\epsilon$.

All Lagrange multipliers are determined, which, again, means that
all constraints are second class. So, every constraint has at
least one non-vanishing Pb with another constraint. The complete
set of constraints is
\begin{eqnarray}
 \begin{array}{ll}
 \theta^0_1=\pi^0_1-M_1B\ , &  \theta_B=\pi_B\ ,  \\
  & \\
 \theta^0_{3/2}=\pi_{3/2}^0\ ,  &   {\theta_{3/2}^0}^\ddag={\pi_{3/2}^0}^\ddag\ ,\\
 \theta_{3/2}^n=\pi_{3/2}^n-\frac{i}{2}\,\psi^{\dagger}_{k}\sigma^{kn}\ , &
       {\theta_{3/2}^n}^\ddag={\pi_{3/2}^n}^\ddag-\frac{i}{2}\,\sigma^{nk}\psi_{k}\ , \\
 \theta_\chi=\pi_{\chi}\ ,  &  \theta_{\chi}^\ddag=\pi_{\chi}^\ddag\ , \\
 \Phi_{3/2}^{0}=-i\sigma^{kn}\partial_k\psi_n-M_{3/2}\left(\gamma^k\psi_k-\chi\right)\ , &
 {\Phi_{3/2}^{0}}^\ddag=-i\partial_k\psi_n^\dagger\sigma^{nk}-M_{3/2}\left(\psi^\dagger_k\gamma^k+\chi^\dagger\right)\ , \\
 \Phi_{\chi}=\gamma^0\psi_0+\gamma^k\psi_k+b\chi\ , &  \Phi_{\chi}^\ddag=-\psi_0^\dagger\gamma^0+\psi_k^\dagger\gamma^k-b\chi^\dagger\ , \\
  & \\
 \theta^{00}_{2}=\pi^{00}_2+\frac{1}{2}\,\partial_nh^{n0}-M_2\eta^0\ , & \theta^{0}_{\eta}=\pi^{0}_{\eta}\ ,\\
 \theta^{0m}_2=\pi^{0m}_2+\partial_nh^{nm}-\frac{1}{2}\,\partial^mh^{00}\qquad  & \theta^{m}_{\eta}=\pi^{m}_{\eta}\ ,\\
 \phantom{\theta^{0m}_2=}-\frac{1}{2}\,\partial^mh^n_n-M_2\eta^m\ , &  \theta_{\epsilon}=\pi_{\epsilon}\ , \\
 \Phi_2^{(2)}=-\partial_n\partial_mh^{nm}+\left(\partial^k\partial_k+M_2^2\right)h^m_m\ , &   \Phi_\eta=h^0_0+h^n_n\ ,\\
 \phantom{\Phi_2^{(2)}=}+2M_2\partial^k\eta_k-2\left(\frac{3+c}{1-c}\right)M_2^2\epsilon\
 , &  \Phi^{(1)}_{2}={\pi_{2}}^{k}_k-\frac{1}{2}\,\partial_nh^{n0}+cM_{2}\eta^0\ .\\
 \end{array}\nonumber\\\label{eqn5.12}
\end{eqnarray}
Again we make linear combinations of constraints in order to
reduce the number of non-vanishing Pb's
\begin{eqnarray}
 \tilde{\Phi}_\chi
&=&
 \Phi_\chi-\frac{b}{M_{3/2}}\,\Phi_{3/2}^0\ ,\nonumber\\
 \tilde{\theta}^n_{3/2}
&=&
 \theta_{3/2}^n-\theta_{3/2}^0\gamma_0\left[(1+b)\gamma^n
 -\frac{b}{M_{3/2}}\,i\overleftarrow{\partial_k}\sigma^{kn}\right]
 +\frac{1}{M_{3/2}}\,\theta_\chi\left[M_{3/2}\gamma^n-i\overleftarrow{\partial_k}\sigma^{kn}\right]
 \ ,\nonumber\\
 \tilde{\Phi}_\chi^\ddag
&=&
 \Phi_\chi^\ddag-\frac{b}{M_{3/2}}\,{\Phi_{3/2}^0}^\ddag
 \ ,\nonumber\\
 \tilde{\theta}_{3/2}^{n\ddag}
&=&
 {\theta_{3/2}^n}^\ddag
 -\left[-(1+b)\gamma^n+\frac{b}{M_{3/2}}\,\sigma^{nk}i\partial_k\right]\gamma_0{\theta_{3/2}^0}^\ddag
 -\frac{1}{M_{3/2}}\left[M_{3/2}\gamma^n-\sigma^{nk}i\partial_k\right]\theta_\chi^\ddag
 \ ,\nonumber\\ \nonumber\\
 \tilde{\Phi}_\eta
&=&
 \Phi_\eta-\frac{1}{M_2}\,\theta^0_\eta\ ,\nonumber\\
 \tilde{\Phi}_2^{(1)}
&=&
 \Phi_2^{(1)}+c\theta_2^{00}+\frac{1}{2M^2}\left(\frac{1-c}{3+c}\right)
 \left(2\partial^k\partial_k+3M^2\right)\theta_\epsilon\ ,\nonumber\\
 \tilde{\theta}^{0n}_{2}
&=&
 \theta^{0n}_{2}+\frac{1}{(3+c)}\,\partial^n\tilde{\Phi}_\eta\ ,\nonumber\\
 \tilde{\Phi}_2^{(2)}
&=&
 \Phi_2^{(2)}+2\partial_k\tilde{\theta}^{0k}_{2}
 \ .\label{eqn5.13}
\end{eqnarray}
With these new constraints the remaining non-vanishing Pb's are
\begin{eqnarray}
 \left\{\theta_1^0(x),\theta_B(y)\right\}_{P}&=&-M_1\delta^3(x-y)\ ,
 \nonumber\\*
   \nonumber\\*
 \left\{\theta^0_{3/2}(x),\tilde{\Phi}_{\chi}(y)\right\}_P&=&\gamma_0\delta^3(x-y) =
 -\left\{{\theta^0_{3/2}}^\ddag(x),\tilde{\Phi}_{\chi}^\ddag(y)\right\}_P\ ,\nonumber\\
 \left\{\theta_{\chi}(x),\Phi^0_{3/2}(y)\right\}_P&=&M_{3/2}\,\delta^3(x-y) =
 -\left\{\theta_{\chi}^\ddag(x),{\Phi^{0}_{3/2}}^\ddag(y)\right\}_P\ ,\nonumber\\
 \left\{\tilde{\theta}^n_{3/2}(x),\tilde{\theta}_{3/2}^{m\ddag}(y)\right\}_P&=&-i\sigma^{mn}\delta^3(x-y)\ , \nonumber\\
   \nonumber\\
 \left\{\theta^{00}_2(x),\theta^0_\eta(y)\right\}_P&=&-M_2\,\delta^3(x-y)\ ,\nonumber\\
 \left\{\tilde{\theta}^{0n}_2(x),\theta^m_\eta(y)\right\}_P&=&-M_2\,g^{nm}\,\delta^3(x-y)\ , \nonumber\\
 \left\{\theta_\epsilon(x),\tilde{\Phi}_2^{(2)}(y)\right\}_P&=&2\left(\frac{3+c}{1-c}\right)M_2^2\,\delta^3(x-y)\ ,\nonumber\\
 \left\{\tilde{\Phi}_2^{(1)}(x),\tilde{\Phi}_\eta(y)\right\}_P&=&-(3+c)\,\delta^3(x-y)\ .
 \label{eqn5.14}
\end{eqnarray}
The Db and the inverse functions that go with them are defined in
\eqref{eqn2.15} and \eqref{eqn2.16}, so we can immediately write
down the ETC and ETAC relations
\begin{eqnarray}
 \left[A^{\mu}(x),\dot{A}^{\nu}(y)\right]_{0}
&=&
 -i\left(g^{\mu\nu}-(1-a)\delta^{\mu}_0\delta^{\nu}_0\right)\delta^3(x-y)
 \ ,\nonumber\\
 \left[A^{\mu}(x),B(y)\right]_{0}
&=&
 \frac{i}{M_1}\,\delta^{\mu}_{0}\delta^3(x-y)\ ,\nonumber\\
 \left[A^{\mu}(x),\dot{B}(y)\right]_{0}
&=&
 -\left[\dot{A}_{\mu}(x),B(y)\right]_{0}=
 -i\delta^{\mu}_k\frac{\partial^k}{M_1}\,\delta^3(x-y)\ ,\nonumber\\
 \left[B(x),\dot{B}(y)\right]_{0}
&=&
 -i\delta^3(x-y)\ ,\label{eqn5.15}
\end{eqnarray}

\begin{eqnarray}
 \left\{\psi^n(x),{\psi^m}^\dagger(y)\right\}_{0}&=&
 -\left[g^{nm}-\frac{1}{2}\,\gamma^{n}\gamma^{m}\right]\delta^3(x-y)\ ,\nonumber\\
 \left\{\psi^0(x),{\psi^0}^\dagger(y)\right\}_{0}&=&
 -\frac{3}{2}\,(1+b)^2\,\delta^3(x-y)\ , \nonumber\\
 \left\{\psi^0(x),{\psi^m}^\dagger(y)\right\}_{0}&=&
 \left[\frac{b+1}{2}\,\gamma^m-b\,\frac{i\partial^m}{M_{3/2}}\right]\gamma_0\,\delta^3(x-y)\ ,\nonumber\\
 \left\{\psi^n(x),{\psi^0}^\dagger(y)\right\}_{0}&=&
 \left[\frac{b+1}{2}\,\gamma^n-b\,\frac{i\partial^n}{M_{3/2}}\right]\gamma_0\,\delta^3(x-y)\ , \nonumber\\
 \left\{\chi(x),\chi^\dagger(y)\right\}_{0}&=&
 -\frac{3}{2}\,\delta^3(x-y)\ , \nonumber\\
 \left\{\psi^0(x),\chi^\dagger(y)\right\}_{0}&=&
 \gamma_0\left[\frac{3(1+b)}{2}-\frac{1}{M_{3/2}}\,i\gamma^k\partial_k\right]\delta^3(x-y)\ ,\nonumber\\
 \left\{\psi^n(x),\chi^\dagger(y)\right\}_{0}&=&
 -\left[\frac{1}{2}\,\gamma^n-\frac{i\partial^n}{M_{3/2}}\right]\delta^3(x-y)\ ,\label{eqn5.16}
\end{eqnarray}
\begin{eqnarray}
 \left[h^{00}(x),\eta^0(y)\right]_0&=&\frac{3}{M_2(3+c)}\,i\delta^3(x-y)\ ,\nonumber\\
 \left[h^{0n}(x),\eta^m(y)\right]_0&=&\frac{1}{M_2}\,g^{nm}\,i\delta^3(x-y)\ ,\nonumber\\
 \left[h^{0n}(x),\epsilon(y)\right]_0&=&-\frac{1}{M_2^2}\left(\frac{1-c}{3+c}\right)
                                        \partial^ni\delta^3(x-y)\ ,\nonumber\\
 \left[h^{nm}(x),\eta^0(y)\right]_0&=&-\frac{1}{M_2(3+c)}\,g^{nm}\,i\delta^3(x-y)\ ,\nonumber\\
 \left[\eta^0(x),\eta^m(y)\right]_0&=&\frac{1}{M_2^2(3+c)}\,\partial^mi\delta^3(x-y)\ ,\nonumber\\
 \left[\eta^0(x),\epsilon(y)\right]_0&=&\frac{3}{2M_2}\frac{(1-c)}{(3+c)^2}\,i\delta^3(x-y)\ .\label{eqn5.17}
\end{eqnarray}
In principle there are also ETC relations among time derivatives
of the fields in \eqref{eqn5.17}, that we have not shown for
convenience. However, they are of importance when calculating the
commutation relations for non-equal times, below.

\subsection{Propagators}\label{prop2}

In order to get commutation and anti-commutation relations for
non-equal times we first construct solutions to the EoMs
(\eqref{eqn4.2}, \eqref{eqn4.3} and \eqref{eqn4.4}) based on the
identities \eqref{eqn3.1}
\begin{eqnarray}
 B(x)
&=&
 \int d^3z\left[\partial_0^z\Delta(x-z;M_B^2)\cdot
 B(z)-\Delta(x-z;M_B^2)\cdot\partial_0^zB(z)\right]\ ,\nonumber\\
 A_\mu(x)
&=&
 \int d^3z\left[\partial_0^z\Delta(x-z;M_1^2)\cdot
 A_\mu(z)-\Delta(x-z;M_1^2)\cdot\partial_0^zA_\mu(z)\right]\nonumber\\
&&
 +\frac{1}{(1-a)M_1^2}\int d^3z\left[\left(\partial_0^z\Delta(x-z;M_B^2)\vphantom{\frac{A}{A}}
 -\partial_0^z\Delta(x-z;M_1^2)\right)\right.\nonumber\\
&&
 \phantom{+\frac{1}{(1-a)M_1^2}\int d^3z[(}\left.\vphantom{\frac{A}{A}}
 -\left(\Delta(x-z;M_B^2)-\Delta(x-z;M_1^2)\right)
 \partial_0^z\right]\nonumber\\
&&
 \phantom{+\frac{1}{(1-a)M_1^2}\int d^3z}
 \times(\Box+M_1^2)A_\mu(z)\ ,\nonumber
\end{eqnarray}
\begin{eqnarray}
 \chi(x)
&=&
 i\int d^3z (i\slpart_x+M_\chi)\gamma^0\Delta(x-z;M^2_{\chi})\chi(z)\ ,\nonumber\\
 \psi_{\mu}(x)
&=&
 i\int d^3z
 (i\slpart_x+M_{3/2})\gamma^0\Delta(x-z;M^2_{3/2})\psi_{\mu}(z)\nonumber\\
&&
 +\frac{2i}{3(b+2)M_{3/2}}\,\int d^3z
 \left[\vphantom{\frac{A}{A}}(i\slpart_x+M_{3/2})\Delta(x-z;M_{3/2}^2)
 \right.\nonumber\\
&&
 \phantom{+\frac{2i}{3(b+2)M_{3/2}}\,\int d^3z\left[\right.}\left.\vphantom{\frac{A}{A}}
 -(i\slpart_x-M_\chi)\Delta(x-z;M^2_{\chi})\right]
 \gamma^0(i\slpart_z-M_{3/2})\psi_{\mu}(z)\nonumber\\
&&
 +\frac{2i}{(3b+2)M_{3/2}}\,\int d^3z
 \left\{\Delta(x-z;M^2_{\chi})
 -\frac{2}{3(b+2)M_{3/2}}\,
 \left[\vphantom{\frac{A}{A}}\right.\right.\nonumber\\
&&
 \left.\left.\hspace{0.8cm}\vphantom{\frac{A}{A}}
 \times(i\slpart_x+M_{3/2})
 \Delta(x-z;M_{3/2}^2)-(i\slpart_x-M_\chi)\Delta(x-z;M^2_{\chi})\right]\right\}\nonumber\\
&&
 \hspace{1cm}\times
 \gamma^0(i\slpart_z+M_\chi)(i\slpart_z-M_{3/2})\psi_{\mu}(z)\ ,\nonumber
\end{eqnarray}
\begin{eqnarray}
 \epsilon(x)
&=&
 \int d^3z\left[\partial^z_0\Delta(x-z;M^2_\epsilon)\cdot\epsilon(z)
 -\Delta(x-z;M^2_\epsilon)\cdot\partial^z_0\epsilon(z)\right]\ ,\nonumber\\
 \eta^\mu(x)
&=&
 \int d^3z\left[\partial^z_0\Delta(x-z;M^2_\eta)\cdot\eta^\mu(z)
 -\Delta(x-z;M^2_\eta)\cdot\partial^z_0\eta^\mu(z)\right]\nonumber\\
&&
 +\frac{1}{M^2_\eta-M^2_\epsilon}\int d^3z
 \left[\partial^z_0\left(\vphantom{\frac{A}{A}}\Delta(x-z;M^2_\epsilon)-\Delta(x-z;M^2_\eta)\right)
 \right.\nonumber\\
&&
 \left.\phantom{+\frac{1}{M^2_\eta-M^2_\epsilon}\int d^3z[}
 -\left(\vphantom{\frac{A}{A}}\Delta(x-z;M^2_\epsilon)-\Delta(x-z;M^2_\eta)\right)
 \cdot\partial^z_0\right]
 (\Box+M^2_\eta)\eta^\mu(z)\ ,\nonumber
\end{eqnarray}
\begin{eqnarray}
 h^{\mu\nu}(x)
&=&
 \int d^3z\left[\partial^z_0\Delta(x-z;M_2^2)\cdot h^{\mu\nu}(z)
 -\Delta(x-z;M_2^2)\cdot\partial^z_0 h^{\mu\nu}(z)\right]\nonumber\\
&&
 +\frac{1}{M_2^2-M^2_\eta}\int d^3z
 \left[\partial^z_0\left(\vphantom{\frac{A}{A}}\Delta(x-z;M^2_\eta)-\Delta(x-z;M_2^2)\right)
 \right.\nonumber\\
&&
 \left.\phantom{+\frac{1}{M^2_\eta-M^2_\epsilon}\int d^3z[}
 -\left(\vphantom{\frac{A}{A}}\Delta(x-z;M^2_\eta)-\Delta(x-z;M_2^2)\right)
 \partial^z_0\right]\nonumber\\
&&
 \phantom{+\frac{1}{M^2_\eta-M^2_\epsilon}\int d^3z}
 \times(\Box+M_2^2)h^{\mu\nu}(z)\nonumber\\
&&
 +\frac{1}{(M^2_\eta-M^2_\epsilon)(M_2^2-M^2_\eta)(M_2^2-M^2_\epsilon)}\int
 d^3z\left[\vphantom{\frac{A}{A}}\right.\nonumber\\
&&
 \partial^z_0\left(\vphantom{\frac{A}{A}}
 (M_2^2-M^2_\eta)\Delta(x-z;M^2_\epsilon)-(M_2^2-M^2_\epsilon)\Delta(x-z;M^2_\eta)
 \right.\nonumber\\
&&
 \phantom{\partial^z_0\left(\right.}\left.\vphantom{\frac{A}{A}}
 +(M^2_\eta-M^2_\epsilon)\Delta(x-z;M_2^2)\vphantom{\frac{A}{A}}\right)\nonumber\\
&&
 -\left(\vphantom{\frac{A}{A}}(M_2^2-M^2_\eta)\Delta(x-z;M^2_\epsilon)
 -(M_2^2-M^2_\epsilon)\Delta(x-z;M^2_\eta)\right.\nonumber\\
&&
 \left.\left.\phantom{-(}
 +(M^2_\eta-M^2_\epsilon)\Delta(x-z;M_2^2)\vphantom{\frac{A}{A}}\right)
 \partial^z_0\right]\left(\Box+M^2_\eta\right)\left(\Box+M_2^2\right)h^{\mu\nu}(z)
 \ .\label{eqn6.1}
\end{eqnarray}
Using these equations \eqref{eqn6.1} and the ETC and ETAC
relations of \eqref{eqn5.15}, \eqref{eqn5.16} and \eqref{eqn5.17}
we obtain the following commutation and anti-commutation relations
\begin{eqnarray}
 \left[B(x),B(y)\right]
&=&
 -i\Delta(x-y,M_B^2)\ ,\nonumber\\
 \left[A^{\mu}(x),B(y)\right]
&=&
 -i\frac{\partial^{\mu}}{M_1}\,\Delta(x-y,M_B^2)\ ,\nonumber\\
 \left[A^{\mu}(x),A^{\nu}(y)\right]
&=&
 -i\left(g^{\mu\nu}+\frac{\partial^{\mu}\partial^{\nu}}{M_1^2}\right)\Delta(x-y;M_1^2)
 +i\frac{\partial^{\mu}\partial^{\nu}}{M_1^2}\Delta(x-y;M_B^2)\nonumber\\
&=&
 P_1^{\mu\nu}i\Delta(x-y;M_1^2)+P_B^{\mu\nu}i\Delta(x-y;M_B^2)\
 ,\label{eqn6.2}
\end{eqnarray}
\begin{eqnarray}
  \left\{\chi(x),\bar{\chi}(y)\right\}
&=&
 -\frac{3}{2}\,i\left(i\slpart+M_\chi\right)\Delta(x-y;M_\chi^2)\ ,\nonumber\\
 \left\{\psi^{\mu}(x),\bar{\chi}(y)\right\}
&=&
 -\frac{1}{2}\left[\gamma^\mu-\frac{2i\partial^\mu}{M_{3/2}}\right]
 i\left(i\slpart+M_\chi\right)\Delta(x-y;M_\chi^2)\
 ,\nonumber\\
 \left\{\psi^{\mu}(x),\bar{\psi}^{\nu}(y)\right\}
&=&
 -i\left(i\slpart+M_{3/2}\right)\left[g^{\mu\nu}-\frac{1}{3}\,\gamma^\mu\gamma^\nu
 +\frac{2\partial^\mu\partial^\nu}{3M_{3/2}^2}\right.\nonumber\\
&&
 \phantom{-i\left(i\slpart+M_{3/2}\right)[}\left.
 -\frac{1}{3M_{3/2}}\left(\gamma^\mu i\partial^\nu-\gamma^\nu i\partial^\mu\right)\right]
 \Delta(x-y;M_{3/2}^2)\nonumber\\
&&
 -\frac{1}{6}\left[\gamma^{\mu}-\frac{2i\partial^{\mu}}{M_{3/2}}\right]
 i\left(i\slpart+M_\chi\right)
 \left[\gamma^{\nu}-\frac{2i\partial^{\nu}}{M_{3/2}}\right]\Delta(x-y;M_\chi^2)\nonumber\\
&=&
 \left(i\slpart+M_{3/2}\right)P_{3/2}^{\mu\nu}i\Delta(x-y;M_{3/2}^2)
 +P_\chi^{\mu\nu}i\Delta(x-y;M_\chi^2)\ ,\label{eqn6.3}
\end{eqnarray}
\begin{eqnarray}
 \left[\epsilon(x),\epsilon(y)\right]
&=&
 -\frac{3}{4}\,\frac{c(1-c)^2}{(3+c)^3}\,i\Delta(x-y;M^2_\epsilon)
 \ ,\nonumber\\
 \left[\eta^\mu(x),\epsilon(y)\right]
&=&
 -\frac{3}{2}\,\frac{(1-c)}{(3+c)^2}\frac{\partial^\mu}{M_2}\,i\Delta(x-y;M^2_\epsilon)\ ,\nonumber\\
 \left[\eta^\mu(x),\eta^\nu(y)\right]
&=&
 \left[g^{\mu\nu}+\frac{\partial^\mu\partial^\nu}{M_\eta^2}\right]i\Delta(x-y;M^2_\eta)
 -\frac{3}{(3+c)}\frac{\partial^\mu\partial^\nu}{M_\eta^2}\,i\Delta(x-y;M^2_\epsilon)\ ,\nonumber\\
 \left[\epsilon(x),h^{\mu\nu}(y)\right]
&=&
 \frac{(1-c)}{(3+c)}
 \left[\frac{\partial^\mu\partial^\nu}{M_2^2}-\frac{1}{2}\,\frac{c}{(3+c)}\ g^{\mu\nu}
 \right]i\Delta(x-y;M^2_\epsilon)\ ,\nonumber\\
 \left[\eta^\alpha(x),h^{\mu\nu}(y)\right]
&=&
 \frac{1}{M_2}\left[\partial^\mu g^{\alpha\nu}+\partial^\nu g^{\alpha\mu}
 +\frac{2}{M_\eta^2}\,\partial^\alpha\partial^\mu\partial^\nu\right]i\Delta(x-y;M^2_\eta)
 \nonumber\\
&&
 -\frac{1}{M_2}\left[\frac{1}{(3+c)}\,\partial^\alpha g^{\mu\nu}
 +\frac{2}{M_\eta^2}\,\partial^\alpha\partial^\mu\partial^\nu\right]i\Delta(x-y;M^2_\epsilon)\ ,\nonumber\\
 \left[h^{\mu\nu}(x),h^{\alpha\beta}(y)\right]
&=&
 \left[g^{\mu\alpha}g^{\nu\beta}+g^{\mu\beta}g^{\nu\alpha}-\frac{2}{3}\,g^{\mu\nu}g^{\alpha\beta}
 \right.\nonumber\\
&&
 +\frac{1}{M_2^2}\left(\partial^\mu\partial^\alpha g^{\nu\beta}+\partial^\nu\partial^\alpha g^{\mu\beta}
 +\partial^\mu\partial^\beta g^{\nu\alpha}+\partial^\nu\partial^\beta g^{\mu\alpha}\right)\nonumber\\
&&
 \left.-\frac{2}{3M_2^2}\left(\partial^\mu\partial^\nu g^{\alpha\beta}+g^{\mu\nu}\partial^\alpha\partial^\beta\right)
 +\frac{4}{3M_2^4}\,\partial^\mu\partial^\nu\partial^\alpha\partial^\beta\right]i\Delta(x-y;M_2^2)\nonumber\\
&&
 -\frac{1}{M_2^2}\left[\partial^\mu\partial^\alpha g^{\nu\beta}+\partial^\nu\partial^\alpha g^{\mu\beta}
 +\partial^\mu\partial^\beta g^{\nu\alpha}+\partial^\nu\partial^\beta
 g^{\mu\alpha}\vphantom{\frac{A}{A}}\right.\nonumber\\
&&
 \left.\phantom{-\frac{1}{M_2^2}[}
 +\frac{4}{M_\eta^2}\,\partial^\mu\partial^\nu\partial^\alpha\partial^\beta\right]
 i\Delta(x-y;M^2_\eta)\nonumber\\
&&
 -\left[\frac{1}{3}\frac{c}{3+c}\ g^{\mu\nu}g^{\alpha\beta}
 -\frac{2}{3M_2^2}\left(\partial^\mu\partial^\nu g^{\alpha\beta}
 +g^{\mu\nu}\partial^\alpha\partial^\beta\right)\right.\nonumber\\
&&
 \left.\phantom{-[}
 +\frac{4(3+c)}{3cM_2^4}\,\partial^\mu\partial^\nu\partial^\alpha\partial^\beta\right]
 i\Delta(x-y;M^2_\epsilon)\nonumber\\
&=&
 2P^{\mu\nu\alpha\beta}_2(\partial)i\Delta(x-y;M_2^2)
 +P^{\mu\nu\alpha\beta}_\eta(\partial)i\Delta(x-y;M^2_\eta)\nonumber\\
&&
 +P^{\mu\nu\alpha\beta}_\epsilon(\partial)i\Delta(x-y;M^2_\epsilon)\ .\label{eqn6.4}
\end{eqnarray}
From the overall minus signs in the (anti-) commutation relations
of the auxiliary fields in \eqref{eqn6.4} we conclude that all
auxiliary fields are ghost, except for the $\epsilon$-field. There
the choice of the gauge parameter $c$ determines whether it is
ghost-like or not: for $-3<c<0$ the $\epsilon$-field is physical
and it ghost-like in all other cases (excluding $c=-3$ and $c=0$).

Having obtained these (anti-) commutation relations we calculate
the propagators
\begin{eqnarray}
 D_{F,a}^{\mu\nu}(x-y)
&=&
 -i<0|T\left[A^{\mu}(x),A^{\nu}(y)\right]|0>\nonumber\\
&=&
 -i\theta(x_0-y_0)\left[P_1^{\mu\nu}(\partial)\Delta^{(+)}(x-y;M_1^2)
 +P_B^{\mu\nu}(\partial)\Delta^{(+)}(x-y;M_B^2)\vphantom{\frac{A}{A}}\right]\nonumber\\
&&
 -i\theta(x_0-y_0)\left[P_1^{\mu\nu}(\partial)\Delta^{(-)}(x-y;M_1^2)
 +P_B^{\mu\nu}(\partial)\Delta^{(-)}(x-y;M_B^2)\vphantom{\frac{A}{A}}\right]\nonumber\\
&=&
 P_1^{\mu\nu}(\partial)\Delta_F(x-y;M_1^2)+P_B^{\mu\nu}(\partial)\Delta_F(x-y;M_B^2)\ .\label{eqn6.5}
\end{eqnarray}
We see that this propagator is explicitly covariant, independent
of the choice of the gauge parameter. Choosing $a=1$ we see that
the terms containing derivatives cancel and that only the
$g^{\mu\nu}$ term remains. It can be seen as the massive photon
propagator. For $a=\infty$ we re-obtain the massive spin-1 field,
like in \eqref{eqn3.3}. Except in the above derivation it is
obtained without non-covariant terms in the propagator. The choice
$a=0$ is particularly interesting, because then still the spin-1
condition $\partial\cdot A=0$ holds (text below \eqref{eqn4.2}),
but the propagator is covariant. In momentum space it looks like
\begin{eqnarray}
 D_{F,0}^{\mu\nu}(P)=\frac{-g^{\mu\nu}+\frac{p^\mu
 p^\nu}{p^2}}{p^2-M_1^2+i\varepsilon}\ .
\end{eqnarray}
The spin-3/2 propagator is
\begin{eqnarray}
 S_{F,b}^{\mu\nu}(x-y)
&=&
 -i<0|T\left[\psi^{\mu}(x),\bar{\psi}^{\nu}(y)\right]|0>\nonumber\\
&=&
 -i\theta(x_0-y_0)\left[\vphantom{\frac{A}{A}}\left(i\slpart+M_{3/2}\right)
 P_{3/2}^{\mu\nu}(\partial)\Delta^{(+)}(x-y;M_{3/2}^2)
 +P_\chi^{\mu\nu}(\partial)\Delta^{(+)}(x-y;M_\chi^2)\vphantom{\frac{A}{A}}\right]\nonumber\\
&&
 -i\theta(x_0-y_0)\left[\vphantom{\frac{A}{A}}\left(i\slpart+M_{3/2}\right)
 P_{3/2}^{\mu\nu}(\partial)\Delta^{(-)}(x-y;M_{3/2}^2)
 +P_\chi^{\mu\nu}(\partial)\Delta^{(-)}(x-y;M_\chi^2)\vphantom{\frac{A}{A}}\right]\nonumber\\
&=&
 \left(i\slpart+M_{3/2}\right)P_{3/2}^{\mu\nu}(\partial)\Delta_F(x-y;M_1^2)+P_\chi^{\mu\nu}(\partial)\Delta_F(x-y;M_B^2)
 \nonumber\\
&&
 +\frac{b}{M_{3/2}}\,\delta_0^\mu\delta_0^\nu\,\delta^4(x-y)\ .\label{eqn6.6}
\end{eqnarray}
Only for $b=0$ we have an explicitly covariant propagator. This
result was also obtained in \cite{babu}. From the text below
\eqref{eqn4.3} we see that the choice $b=0$ means that we have
only one of the two spin-3/2 conditions or, to put it in a
different way, we have added an extra spin-1/2 piece to make the
RS propagator explicitly covariant.

For $b=-\frac{4}{3}$ and $b=-1$ we have that
$i\partial\cdot\psi=0$ (, but $\gamma\cdot\psi\neq0$), but then
the propagator is not covariant anymore.

The spin-2 propagator is
\begin{eqnarray}
 D_{F,c}^{\mu\nu\alpha\beta}(x-y)
&=&
 -i<0|T\left[h^{\mu\nu}(x)h^{\alpha\beta}(y)\right]|0>\nonumber\\
&=&
 -i\theta(x^0-y^0)\left[
 2P^{\mu\nu\alpha\beta}_2(\partial)\Delta^{(+)}(x-y;M^2)
 +P^{\mu\nu\alpha\beta}_\eta(\partial)i\Delta^{(+)}(x-y;M^2_\eta)\right.\nonumber\\
&&
 \left.\phantom{-i\theta(x^0-y^0)[}
 +P^{\mu\nu\alpha\beta}_\epsilon(\partial)i\Delta^{(+)}(x-y;M^2_\epsilon)\right]\nonumber\\
&&
 -i\theta(y^0-x^0)\left[
 2P^{\mu\nu\alpha\beta}_2(\partial)\Delta^{(-)}(x-y;M^2)
 +P^{\mu\nu\alpha\beta}_\eta(\partial)i\Delta^{(-)}(x-y;M^2_\eta)\right.\nonumber\\
&&
 \left.\phantom{-i\theta(y^0-x^0)[}
 +P^{\mu\nu\alpha\beta}_\epsilon(\partial)i\Delta^{(-)}(x-y;M^2_\epsilon)\right]\nonumber\\
&=&
 2P^{\mu\nu\alpha\beta}_2(\partial)\Delta_F(x-y;M^2)
 +P^{\mu\nu\alpha\beta}_\eta(\partial)\Delta_F(x-y;M^2_\eta)\nonumber\\
&&
 +P^{\mu\nu\alpha\beta}_\epsilon(\partial)\Delta_F(x-y;M^2_\epsilon)\
 .\label{eqn6.7}
\end{eqnarray}
We see that this propagator \eqref{eqn6.7} does not contain local,
non-covariant terms independent of the choice of the gauge
parameter. The first part of \eqref{eqn6.7}
($P^{\mu\nu\alpha\beta}_2(\partial)$-part) is pure spin-2
\footnote{The factor $2$ can again be transformed away by
redefining all fields as in \eqref{eqn3.2}}. The nature of the
other parts depends on the free gauge parameter.

Since $c$ is still a free parameter it is interesting to look at
several gauges. But before that, we exclude $c=1$ and $c=-3$ as
before. In these cases the $\epsilon$-field vanishes and the EoM
are quite different. Also the quantization procedure runs
differently.

An interesting gauge which we want to discuss here is $c=-1$. From
(\ref{eqn4.4}) we see that all fields become free Klein-Gordon
fields of mass $M_2$. As a result of this choice all derivative
terms disappear in (\ref{eqn6.7}) and what is left is
\begin{eqnarray}
 D_{F,-1}^{\mu\nu\alpha\beta}(x-y)
&=&
 \left[g^{\mu\alpha}g^{\nu\beta}+g^{\mu\beta}g^{\nu\alpha}-\frac{1}{2}\,g^{\mu\nu}g^{\alpha\beta}
 \right]\Delta_F(x-y;M^2)\ .\qquad\qquad\label{eqn6.8}
\end{eqnarray}
In contrast to the spin-1 case, discussed above, equation
\eqref{eqn6.8} is not the massive version of the massless spin-2
propagator.

Equation \eqref{eqn6.7} yields for the choice $c=0$
\begin{eqnarray}
 D_{F,0}^{\mu\nu\alpha\beta}(x-y)
&=&
 2P^{\mu\nu\alpha\beta}_2(\partial)\Delta_F(x-y;M_2^2)
 -\frac{1}{M_2^2}\left[\vphantom{\frac{A}{A}}\partial^\mu\partial^\alpha g^{\nu\beta}
 +\partial^\nu\partial^\alpha g^{\mu\beta}
 +\partial^\mu\partial^\beta g^{\nu\alpha}
 +\partial^\nu\partial^\beta g^{\mu\alpha}
 \right.\nonumber\\
&&
 \left.-\frac{2}{3}\left(\partial^\mu\partial^\nu g^{\alpha\beta}
 +g^{\mu\nu}\partial^\alpha\partial^\beta\right)
 +\frac{4\partial^\mu\partial^\nu\partial^\alpha\partial^\beta}{3M_2^2}\right]\Delta_F(x-y)
 \nonumber\\
&&
 +\frac{4}{3M_2^2}\,\partial^\mu\partial^\nu\partial^\alpha\partial^\beta
 \tilde{\Delta}_F(x-y)\ ,\nonumber\\
 D_{F,0}^{\mu\nu\alpha\beta}(p)
&=&
 \left[g^{\mu\alpha}g^{\nu\beta}+g^{\mu\beta}g^{\nu\alpha}-\frac{2}{3}\,g^{\mu\nu}g^{\alpha\beta}
 +\frac{2}{3p^2}\left(p^\mu p^\nu g^{\alpha\beta}+g^{\mu\nu}p^\alpha p^\beta\right)\right.\nonumber\\
&&
 \left.-\frac{1}{p^2}\left(p^\mu p^\alpha g^{\nu\beta}+p^\nu p^\alpha g^{\mu\beta}
 +p^\mu p^\beta g^{\nu\alpha}+p^\nu p^\beta g^{\mu\alpha}\right)
 +\frac{4}{3p^4}\,p^\mu p^\nu p^\alpha
 p^\beta\right]\nonumber\\
&&
 \times\frac{1}{p^2-M_2^2+i\varepsilon}\ .\label{eqn6.10}
\end{eqnarray}
Here, the $\tilde{\Delta}_F(x-y)$ (as well as various other
$\Delta$ propagators) is defined in appendix \ref{deltaprop}. As
in the spin-1 case this propagator \eqref{eqn6.10} satisfies the
field equations (and is therefore pure spin-2) and is explicitly
covariant. This result is also obtained by ignoring the $c$ term
in the Lagrangian (\ref{eqn4.1}) from the outset.

\subsection{Massless limit}\label{massless}

It is most easy to study the massless limits of the propagators
obtained in the previous subsection in momentum space
\begin{eqnarray}
 \underset{M_{1}\rightarrow0}{Lim}D_{F,a}^{\mu\nu}(p)
&=&
 \left[-g^{\mu\nu}+\left(1-a\right)\frac{p^\mu p^\nu}{p^2}\right]\frac{1}{p^2+i\varepsilon}\
 .\label{eqn7.1}
\end{eqnarray}
Although we have not presented the massless case, it is done
rather easily. The quantization procedure runs very similar to
what is presented in section \ref{quant2}, contrary to the case
without an auxiliary field (section \ref{quant1}), only the
equations like in \eqref{eqn6.1} are a bit different. It should be
noticed that it is sufficient in the massless case to ignore the
mass term of the spin-1 field in \eqref{eqn4.1a}, only. So, even
though allowing for a mass term for the auxiliary field, both
$A^\mu$ and $B$ turn out to be massless. Therefore the freedom in
choosing the gauge parameter is still present. In the massless
case the exact same result as \eqref{eqn7.1} is obtained, so the
massless limit connects smoothly with the massless case and is
explicitly covariant. In fact this line of reasoning is valid for
all three spin cases with auxiliary fields. Having mentioned this,
we will not come back to this when discussing the massless limits
of the spin-3/2 and spin-2 cases below.

The massless limit of the spin-3/2 field is
\begin{eqnarray}
 \underset{M_{3/2}\rightarrow0}{Lim}S_{F,0}^{\mu\nu}(p)
&=&
 -\slp\left[g^{\mu\nu}-\frac{1}{2}\,\gamma^\mu\gamma^\nu\right]\frac{1}{p^2+i\varepsilon}
 +\gamma^\mu p^\nu\frac{1}{p^2+i\varepsilon}-2p^\mu p^\nu\slp\ \frac{1}{p^4+i\varepsilon}
 \ .\label{eqn7.2}
\end{eqnarray}
We notice that when this propagator (\ref{eqn7.2}) is coupled to
conserved currents only the first two parts contribute. These
parts form exactly the massless spin-3/2 propagator with only the
helicities $\lambda=\pm 3/2$ (\cite{deser}). When we couple the
(massive) RS-propagator \eqref{eqn3.4} to conserved currents and
take the massless limit \footnote{Terms in the massive RS
propagator that do not have a proper massless limit do not
contribute since we couple to conserved currents} we see that it
is different from the one in \eqref{eqn7.2} because of the factor
in front of the $\gamma^\mu\gamma^\nu$ term.

The massless limit of the spin-2 propagator is
\begin{eqnarray}
 \underset{M_{2}\rightarrow0}{Lim}D_{F,c}^{\mu\nu\alpha\beta}(p)
&=&
 \left[g^{\mu\alpha}g^{\nu\beta}+g^{\mu\beta}g^{\nu\alpha}-\frac{2+c}{3+c}\,g^{\mu\nu}g^{\alpha\beta}
 \right]\frac{1}{p^2+i\varepsilon}\nonumber\\
&&
 -(1+c)\frac{1}{p^2}\left[\vphantom{\frac{A}{A}} p^\mu p^\alpha g^{\nu\beta}+p^\nu p^\alpha g^{\mu\beta}
 +p^\mu p^\beta g^{\nu\alpha}+p^\nu p^\beta g^{\mu\alpha}\right.\nonumber\\
&&
 \left.\phantom{-(1+c)\frac{1}{p^2}(}
 -\frac{2}{3+c}\left(p^\mu p^\nu g^{\alpha\beta}+g^{\mu\nu}p^\alpha p^\beta\right)\right]
 \frac{1}{p^2+i\varepsilon}\nonumber\\
&&
 +\frac{4(1+c)^2}{3+c}\,\frac{p^\mu p^\nu p^\alpha
 p^\beta}{p^4}\frac{1}{p^2+i\varepsilon}\ .\label{eqn7.3}
\end{eqnarray}
Making the choice of the gauge parameter $c\rightarrow\pm\infty$
we see that \eqref{eqn7.3} becomes the massless spin-2 propagator
plus terms proportional to $p$. In physical processes these terms
do not contribute when coupled to conserved currents
\begin{eqnarray}
 D^{\mu\nu\alpha\beta}_{F,\pm\infty}(p)
&=&
 \left[g^{\mu\alpha}g^{\nu\beta}+g^{\mu\beta}g^{\nu\alpha}-\,g^{\mu\nu}g^{\alpha\beta}
 \right]\frac{1}{p^2+i\varepsilon}+O(p)\ .\qquad\label{eqn7.4}
\end{eqnarray}
Again, this is different from taking the massive spin-2 propagator
\eqref{eqn3.5}, couple it to conserved currents and taking the
massless limit, as is mentioned in \cite{veltman}.

Having obtained the correct massless spin-2 propagator
\eqref{eqn7.3} it is particularly interesting to see how this
limit comes about. Considering the propagator \eqref{eqn6.7}
(coupled to conserved currents) with a small non-zero mass and
requiring that it is a mixture of pure spin-2 and spin-0 (so no
ghosts or tachyons) in order to have a kind of massive Brans-Dicke
\cite{brans} theory, this would imply that $-3<c<0$. However with
this restriction we cannot take the mass smoothly to zero in order
to have a pure massless spin-2 propagator, because this requires
$c\rightarrow\pm\infty$ as mentioned before.

The above situation of a pure massive spin-2 and spin-0 propagator
limiting smoothly to a pure massless spin-2 propagator can be
obtained in \cite{kimura}, but there the set-up is quite different
as well as the original goal.

\subsection{Momentum Representation}\label{momrep}

To finalize the description of the higher spin fields coupled to
auxiliary fields we give the momentum representation of these
fields in this subsection. Also, we give the relations which hold
for the various creation and annihilation operators.

A solution to the EoM of the fields in \eqref{eqn4.2},
\eqref{eqn4.3} and \eqref{eqn4.4} in terms of the auxiliary fields
is
\begin{eqnarray}
 A_\mu
&=&
 V_{\mu}+\frac{\partial_\mu}{M_1}\,B\ ,\nonumber\\
 \psi_{\mu}
&=&
 \Psi_{\mu}+\frac{1}{3}\left(\gamma_{\mu}-\frac{2i\partial_{\mu}}{M_{3/2}}\right)\chi\
 ,\nonumber\\
 \eta_{\mu}
&=&
 \Phi_{1,\mu}+\frac{2(3+c)}{c(1-c)}\,\frac{\partial_\mu}{M_2}\,\epsilon\ ,\nonumber\\
 h_{\mu\nu}
&=&
 \Phi_{2,\mu\nu}-\frac{1}{M_2}\left(\partial_\mu\Phi_{1,\nu}+\partial_\nu\Phi_{1,\mu}\right)
 +\frac{2}{3}\,\frac{3+c}{1-c}\left(g_{\mu\nu}-\frac{2(3+c)}{c}\,\frac{\partial_\mu\partial_\nu}{M_2^2}\right)
 \epsilon\ ,\label{eqn8.1}
\end{eqnarray}
where
\begin{eqnarray}
 (\Box+M_1^2)V_\mu=0\quad , & \quad\partial\cdot V=0\quad , & \nonumber\\
 (i\slpart-M_{3/2})\Psi_\mu=0\quad , & \quad\gamma\cdot\Psi=0\quad , & \quad i\partial\cdot\Psi=0\ ,\nonumber\\
 (\Box+M_2^2)\Phi_{2,\mu\nu}=0\quad , & \quad\partial^{\mu}\Phi_{2,\mu\nu}=0\quad , & \quad
 \Phi^{\mu}_{2,\mu}=0\ ,\label{eqn8.2}
\end{eqnarray}
and are therefore free spin-1, spin-3/2 and spin-2 fields,
respectively. The field $\Phi_{1,\mu}$ also satisfies the free
spin-1 equations, but is of negative norm as we will see below.

Since the anti-commutator of the $\chi$-field \eqref{eqn6.3} and
the commutator of the $\epsilon$-field \eqref{eqn6.4} contain
constants we redefine these fields for convenience
\begin{eqnarray}
 \chi&=&\sqrt{\frac{3}{2}}\,\chi'\ \nonumber\\
 \epsilon&=&\frac{\sqrt{3}(1-c)}{2(3+c)}\ \epsilon'\ .\label{eqn8.3}
\end{eqnarray}
\footnote{The part in the commutator of the $\epsilon$-field that
determines whether $\epsilon$ is ghost-like or not is not taken in
the redefinition.} Therefore \eqref{eqn8.1} becomes
\begin{eqnarray}
 \psi_{\mu}
&=&
 \Psi_{\mu}+\frac{1}{\sqrt{6}}\left(\gamma_{\mu}-\frac{2i\partial_{\mu}}{M_{3/2}}\right)\chi'\
 ,\nonumber\\
 \eta_{\mu}
&=&
 \Phi_{1,\mu}+\frac{\sqrt{3}}{c}\,\frac{\partial_\mu}{M_2}\,\epsilon'\ ,\nonumber\\
 h_{\mu\nu}
&=&
 \Phi_{2,\mu\nu}-\frac{1}{M_2}\left(\partial_\mu\Phi_{1,\nu}+\partial_\nu\Phi_{1,\mu}\right)
 +\frac{1}{\sqrt{3}}\left(g_{\mu\nu}-\frac{2(3+c)}{c}\,\frac{\partial_\mu\partial_\nu}{M_2^2}\right)
 \epsilon'\ .\label{eqn8.4}
\end{eqnarray}
The momentum representation of the fields is
\begin{eqnarray}
 B(x)
&=&
 \int\frac{d^3p}{(2\pi)^32E_B}\left[a_B(p)e^{-ipx}+a_B^\dagger(p)e^{ipx}\right]_{p^0=E_B}\ ,\nonumber\\
 V_{\mu}(x)
&=&
 \sum_{\lambda=-1}^{1}\int\frac{d^3p}{(2\pi)^32E_V}\left[a_{V,\mu}(p\lambda)e^{-ipx}+a^\dagger_{V,\mu}(p\lambda)e^{ipx}\right]
 _{p^0=E_V}\ ,\nonumber
\end{eqnarray}
\begin{eqnarray}
 \chi'(x)
&=&
 \sum_{s=-\frac{1}{2}}^{\frac{1}{2}}\int\frac{d^3p}{(2\pi)^32E_{\chi}}\left[b_\chi(ps)u_\chi(ps)e^{-ipx}
 +d^\dagger_\chi(ps)v_\chi(ps)e^{ipx}\right]_{p^0=E_{\chi}}\ ,\nonumber\\
 \Psi_{\mu}(x)
&=&
 \sum_{s=-\frac{3}{2}}^{\frac{3}{2}}\int\frac{d^3p}{(2\pi)^32E_\Psi}\left[b_{\Psi}(ps)u_\mu(ps)e^{-ipx}
 +d^\dagger_{\Psi}(ps)v_\mu(ps)e^{ipx}\right]_{p^0=E_\Psi}\
 ,\nonumber
\end{eqnarray}
\begin{eqnarray}
 \epsilon'(x)
&=&
 \int\frac{d^3p}{(2\pi)^32E_{\epsilon}}\left[a_{\epsilon}(p)e^{-ipx}+a^\dagger_{\epsilon}(p)e^{ipx}\right]_{p^0=E_{\epsilon}}
 \ ,\nonumber\\
 \Phi_{1,\mu}(x)
&=&
 \sum_{\lambda=-1}^{1}\int\frac{d^3p}{(2\pi)^32E_1}\left[a_{1,\mu}(p\lambda)e^{-ipx}
 +a^\dagger_{1,\mu}(p\lambda)e^{ipx}\right]_{p^0=E_1}
 \ ,\nonumber\\
 \Phi_{2,\mu\nu}
&=&
 \sum_{\lambda=-2}^{2}\int\frac{d^3p}{(2\pi)^32E_2}\left[a_{2,\mu\nu}(p\lambda)e^{-ipx}
 +a^\dagger_{2,\mu\nu}(p\lambda)e^{ipx}\right]_{p^0=E_2}
 \ ,\qquad\quad\label{eqn8.5}
\end{eqnarray}
where $E_i=\sqrt{|\vec{p}|^2+M_i^2}$. In \eqref{eqn8.5} the
spin-3/2 spinor $u_\mu(ps)$ is a tensor product of a spin-1
polarization vector and a spin-1/2 spinor:
$u_\mu=\epsilon_\mu\otimes u$. The normalization of this
(spin-1/2) spinor, as well as that of $u_\chi$, is
$\bar{u}(ps)u(ps')=2M\delta_{ss'}$ and of course something similar
for the $v$-spinors. With this normalization the creation and
annihilation operators satisfy the following (commutation)
relations
\begin{eqnarray}
 \left[a_{B}(p),a_{B}^\dagger(p')\right]
&=&
 -(2\pi)^32E_B\,\delta^3(p-p')\ ,\nonumber\\
 \left[a_{V,\mu}(p\lambda),a_{V,\nu}^\dagger(p'\lambda')\right]
&=&
 \left(-g_{\mu\nu}+\frac{p_\mu p_\nu}{M_1^2}\right)(2\pi)^32E_V\,\delta^3(p-p')\delta_{\lambda\lambda'}\ ,\nonumber\\
 &&\nonumber\\
 \left\{b_\chi(ps),b_\chi^\dagger(p's')\right\}
&=&
 \left\{d_\chi(ps),d_\chi^\dagger(p's')\right\}
 =-(2\pi)^32E_\chi\,\delta^3(p-p')\delta_{ss'}\ ,\nonumber\\
 \left\{b_{\Psi}(ps),b_{\Psi}^\dagger(p's')\right\}
&=&
 \left\{d_{\Psi}(ps),d_{\Psi}^\dagger(p's')\right\}
 =(2\pi)^32E_\Psi\,\delta^3(p-p')\delta_{ss'}\
 ,\nonumber
\end{eqnarray}
\begin{eqnarray}
 \left[a_{\epsilon}(p),a_{\epsilon}^\dagger(p')\right]
&=&
 -\frac{c}{3+c}(2\pi)^32E_{\epsilon}\,\delta^3(p-p')\ ,\nonumber\\
 \left[a_{1,\mu}(p\lambda),a_{1,\nu}^\dagger(p'\lambda')\right]
&=&
 -\left(-g_{\mu\nu}+\frac{p_{\mu}p_\nu}{M_\eta^2}\right)(2\pi)^32E_1\,\delta^3(p-p')\delta_{\lambda\lambda'}\ ,\nonumber\\
 \left[a_{2,\mu\nu}(p\lambda),a_{2,\alpha\beta}(p'\lambda')\right]
&=&
 \left[g_{\mu\alpha}g_{\nu\beta}+g_{\mu\beta}g_{\nu\alpha}-\frac{2}{3}\,g_{\mu\nu}g_{\alpha\beta}
 \right.\nonumber\\
&&
 \ -\frac{1}{M_2^2}\left(p_\mu p_\alpha g_{\nu\beta}+p_\nu p_\alpha g_{\mu\beta}
 +p_\mu p_\beta g_{\nu\alpha}+p_\nu p_\beta g_{\mu\alpha}\right)\nonumber\\
&&
 \left.\ +\frac{2}{3M_2^2}\left(p_\mu p_\nu g_{\alpha\beta}+g_{\mu\nu}p_\alpha p_\beta\right)
 +\frac{4}{3M_2^4}\,p_\mu p_\nu p_\alpha p_\beta\right]\nonumber\\
&&
 \times(2\pi)^32E_2\,\delta^3(p-p')\delta_{\lambda\lambda'}\ .\label{eqn8.6}
\end{eqnarray}
All other (anti-) commutation relations vanish. These (anti-)
commutation relations are such that the relations in
\eqref{eqn6.2}, \eqref{eqn6.3} and \eqref{eqn6.4} remain valid.

To complete the properties of the fields in momentum space there
still are the following relations
\begin{eqnarray}
 p^{\mu}a_{V,\mu}(p\lambda)=0\ ,&&\nonumber\\*
 \nonumber\\*
 p^{\mu}u_{\mu}(ps)=0\ ,&& \gamma^{\mu}u_{\mu}(ps)=0\
 ,\nonumber\\*
 \nonumber\\*
 p^{\mu}a_{1,\mu}(p\lambda)=0\ ,&&\nonumber\\
 p^{\mu}a_{2,\mu\nu}(p\lambda)=0\ ,&& a_{2,\mu}^{\mu}(p\lambda)=0\
 .\label{eqn8.7}
\end{eqnarray}

\section{Conclusion and Discussion}

We conclude this article by stating that we have quantized
(massive) higher spin ($1\leq j\leq 2$) fields in both the case
where they are free (section \ref{ffields}) and where they are
coupled to (an) auxiliary field(s) (section \ref{aux}). We have
presented a full constraint analysis and quantization procedure to
come to equal time (anti) commutation relations.

In the free case we have explicitly shown that the constructed
propagators are non-covariant, which is well known. In the coupled
case, i.e. auxiliary fields are coupled to gauge conditions of the
free case, the propagators can be covariant. Only in the spin-3/2
case this requires a choice of the parameter, namely $b=0$. The
obtained propagators have a smooth massless limit and connect
perfectly to propagators which would be obtained in the massless
case (including (an) auxiliary field(s)).

When coupled to conserved currents we see that it is possible to
obtain the correct massless spin-$j$ propagators carrying only the
helicities $\lambda=\pm j_z$. Only in the spin-3/2 and in the
spin-2 case we have to make choices for the parameters, namely
$b=0$ and $c=\pm\infty$. As far as these two cases is concerned,
it is a different situation then taking the massive propagator,
couple it to conserved currents and putting the mass to zero. We
stress that in these cases the limits are only smooth if the
massive propagators contain ghost parts.

\begin{appendices}
\section{$\Delta$ Propagators}\label{deltaprop}

A few definitions of on mass-shell propagators, according to
\cite{Bjorken}, are
\begin{eqnarray}
 \Delta(x;m^2)&=&\frac{-i}{(2\pi)^3}\int d^4p\epsilon(p_0)\delta(p^2-m^2)e^{-ipx}\ ,
 \nonumber\\
 \Delta^{\pm}(x;m^2)&=&(2\pi)^{-3}\int d^4p\theta(\pm p_0)\delta(p^2-m^2)e^{-ipx}\ ,
 \nonumber\\
 \Delta^{(1)}(x;m^2)&=&\frac{1}{(2\pi)^3}\int d^4p\,\delta(p^2-m^2)e^{-ipx}\ , \label{e1.1}
\end{eqnarray}
which satisfy the relations amongst each other
\begin{eqnarray}
 i\Delta(x;m^2)&=&\Delta^{+}(x;m^2)-\Delta^{-}(x;m^2)\ ,\nonumber\\
 \Delta^{+}(-x;m^2)&=&\Delta^{-}(x;m^2)\ ,\nonumber\\
 \Delta^{(1)}(x;m^2)&=&\Delta^{+}(x;m^2)+\Delta^{-}(x;m^2)\
 .\label{e1.1a}
\end{eqnarray}
Furthermore, there are the following Green functions
\begin{eqnarray}
 -\Delta_F(x;m^2)&=&i\left[\theta(x_0)\Delta^{+}(x;m^2)+\theta(-x_0)\Delta^{-}(x;m^2)\right]\ ,\nonumber\\
 \Delta_{ret}(x;m^2)&=&-\theta(x^0)\Delta(x;m^2)\ ,\nonumber\\
 \Delta_{adv}(x;m^2)&=&\theta(-x^0)\Delta(x;m^2)\ ,\nonumber\\
 \bar{\Delta}(x;m^2)&=&-\frac{1}{2}\,\epsilon(x-y)\Delta(x;m^2)\ ,\label{e1.1b}
\end{eqnarray}
where the Green function of the last line of \eqref{e1.1b} is
defined in the book of Nakanishi and Ojima (see \cite{Nakan2}). A
well known form the the Feynman propagator $\Delta_F(x-y)$ is
\begin{eqnarray}
  \Delta_F(x;m^2)&=&\frac{1}{(2\pi)^4}\int d^4p\ \frac{e^{-ipx}}{p^2-m^2+i\varepsilon}\
  .
\end{eqnarray}
The following $\Delta$ propagators are defined to be
\begin{eqnarray}
 \tilde{\Delta}(x)&=&-\frac{\partial}{\partial m^2}\,\Delta(x;m^2)|_{m^2=0}\ ,\nonumber\\
 \tilde{\tilde{\Delta}}(x)&=&\left(\frac{\partial}{\partial m^2}\right)^2\Delta(x;m^2)|_{m^2=0}
 \ .\label{e1.2}
\end{eqnarray}
Since the last two lines of \eqref{e1.2} are also valid for
Feynman function we can, by using the integral representation of
the Feynman function \eqref{e1.1b} give integral representations
for $\tilde{\Delta}_F(x)$ and $\tilde{\tilde{\Delta}}_F(x)$
\begin{eqnarray}
 \tilde{\Delta}_F(x;m^2)
&=&
 -\frac{1}{(2\pi)^4}\int d^4p\ \frac{e^{-ipx}}{p^4+i\varepsilon}\
 ,\nonumber\\*
 \tilde{\tilde{\Delta}}_F(x;m^2)
&=&
 \frac{1}{(2\pi)^4}\int d^4p\ \frac{e^{-ipx}}{p^6+i\varepsilon}\
 .\label{e1.2a}
\end{eqnarray}
Furthermore we have the important relations
\begin{eqnarray}
 \left(\Box+m^2\right)\Delta(x;m^2)&=&0\ ,\nonumber\\
 \Delta(x;m^2)|_0&=&0\ ,\nonumber\\
 \left[\partial_0\Delta(x;m^2)\right]|_0&=&-\delta(\vec{x})\
 ,\nonumber
\end{eqnarray}
\begin{eqnarray}
 \Box\tilde{\Delta}(x)&=&\Delta(x)\ ,\nonumber\\
 \tilde{\Delta}(x)|_{0}&=&\partial_0\tilde{\Delta}(x)|_{0}
 =\partial_0^2\tilde{\Delta}(x)|_{0}=0\
 ,\nonumber\\
 \partial_0^3\tilde{\Delta}(x)|_{0}&=&-\delta(\vec{x})\ ,\nonumber
\end{eqnarray}
\begin{eqnarray}
 \Box\tilde{\tilde{\Delta}}(x)&=&\tilde{\Delta}(x)\ ,\nonumber\\
 \tilde{\tilde{\Delta}}(x)|_{0}&=&\partial_0\tilde{\tilde{\Delta}}(x)|_{0}
 =\ldots=\partial_0^4\tilde{\tilde{\Delta}}(x)|_{0}=0\
 ,\nonumber\\
 \partial_0^5\tilde{\tilde{\Delta}}(x)|_{0}&=&-\delta(\vec{x})\ ,\nonumber
\end{eqnarray}
\begin{eqnarray}
 \left[\partial_0\Delta^{(1)}(x;m^2)\right]|_0&=&0\ .
\label{e1.3}
\end{eqnarray}
\end{appendices}

\end{document}